\documentclass[fleqn,usenatbib,useAMS]{mnras}

\usepackage{graphicx}	
\usepackage{amsmath}	
\usepackage{amssymb}	
\usepackage{bm}		
\usepackage{pdflscape}	

\newcommand{\gad}{{\sc Gadget-3}}
\newcommand{\gizmo}{{\sc Gizmo}}

\newcommand{\simba}{{\sc Simba}}
\newcommand{\martini}{{\sc Martini}}
\newcommand{\caesar}{{\sc Caesar}}

\newcommand{\hmpc}{\,h^{-1}{\rm Mpc}}
\newcommand{\HI}{\ion{H}{i}}

\usepackage[T1]{fontenc}
\usepackage{ae,aecompl}

\usepackage{graphicx}
\usepackage{epstopdf}
\usepackage{subcaption}
\usepackage{caption}
\usepackage{gensymb}
\usepackage{calc}
\usepackage{pdflscape}
\usepackage{rotating}
\usepackage{afterpage}

\title[]{The redshift evolution of the baryonic Tully-Fisher relation in \simba\ }

\author[]{M.~Glowacki$^{1,2}${\thanks{Contact e-mail: \href{mailto:marcin@idia.ac.za}{marcin@idia.ac.za}}, E.~Elson$^{2}$, R.~Dav\'e$^{3,2,4}$}\\
$^{1}$Inter-University Institute for Data Intensive Astronomy, Bellville 7535, South Africa\\
$^{2}$Department of Physics and Astronomy, University of the Western Cape, Robert Sobukwe Road, Bellville 7535, South Africa\\
$^{3}$Institute for Astronomy, Royal Observatory, Univ. of Edinburgh, Edinburgh EH9 3HJ, UK\\
$^{4}$South African Astronomical Observatories, Observatory, Cape Town 7925, South Africa
}
\date{}

\pubyear{2019}

\hypersetup{draft}

\begin{document}
\label{firstpage}
\pagerange{\pageref{firstpage}--\pageref{lastpage}}
\maketitle

\begin{abstract}
The baryonic Tully-Fisher relation (BTFR) is an important tool for constraining galaxy evolution models. As 21-cm H{\sc i} emission studies have been largely restricted to low redshifts, the redshift evolution of the BTFR is less studied. The upcoming LADUMA survey (Looking At the Distant Universe with the MeerKAT Array) will address this. As preparation for LADUMA, we use the \simba\ hydrodynamical galaxy formation simulation from the \simba-hires $(25\hmpc)^3$ run to generate rotational velocity measures from galaxy rotation curves ($V_{\rm flat}$) and H{\sc i} spectral line profile widths ($W_{\rm 50}$ and $W_{\rm 20}$) at three different redshifts ($z$~=~0, 0.5, and 1). Using these measures, together with the dark matter velocity dispersion and halo mass, we consider the redshift evolution of the BTFR of \simba\ galaxies. We find that LADUMA will be successful in detecting weak redshift evolution of the BTFR, provided that auxiliary data is used to distinguish galaxies with disky morphologies. $W_{\rm 20}$ spectral line widths give lower scatter and more pronounced redshift evolution compared to $W_{\rm 50}$. We also compare these rotational velocity measures to the dark matter velocity dispersion across redshift and galaxy morphology. We find weak redshift evolution between rotational velocity and the dark matter halo mass, and provide fits for estimating a galaxy's dark matter halo mass from H{\sc i} spectral line widths. This study with \simba\ showcases the importance of upcoming, deep SKA pathfinder surveys such as LADUMA, and provides predictions to compare with redshift evolution of the BTFR and galaxy dark matter content from H{\sc i} rotational velocity measures. 
\end{abstract}





\section{Introduction}\label{sec:intro}

The relation between the rotation speeds of galaxies and their luminosity, the Tully-Fisher relation \cite[TFR;][]{Tully1977}, gives us an important constraint on galaxy assembly history and feedback processes~\cite[e.g.][]{Haynes1999,Sanders2002,Springob2007,deRossi2010,Ponomareva2018}. For instance, an even tighter version of this relation spanning multiple orders of magnitude, which includes the gas and stars of galaxies (McGaugh 2012), provides evidence for a link between dark matter halos and the baryonic content of the galaxies associated with them. This relation is called the baryonic Tully-Fisher relation (henceforth BTFR).

There are two main methods in observational studies of the BTFR for obtaining the rotational velocity of galaxies through the neutral hydrogen (H{\sc i}) gas content traced by the 21-cm `spin-flip' transition. The first applies primarily to local universe studies, where it is viable to spatially resolve the H{\sc i} content and derive a rotation curve for the galaxy, which in turn provides a rotational velocity depending on one's method of velocity selection. At higher redshifts, current telescopes cannot image galaxies at the spatial resolution required to measure rotation curves. Alternatively, one can use the spectral line widths of the H{\sc i} emission profile as a measure for the rotational velocity to study the BTFR. Both methods have been used extensively \cite[e.g.][]{Verheijen2001,Noordermeer2007,Gurovich2010,McGaugh2012,Zaritsky2014,Lelli2016,Ponomareva2017}. 

The BTFR found from rotation curves and spectral line widths will vary, as demonstrated in \cite{Ponomareva2018} and \cite{Lelli2019}. It is thus important to ask, how does using the H{\sc i} 21\,cm spectral line widths compare to the rotation curve velocity values? Moreover, how do these two methods compare to the true velocity dispersion from the dark matter halo component of these galaxies, which the BTFR aims to trace? This is not something that can be easily answered from observations, given the inability to directly observe the dark matter content of galaxies.
While a tight relation has been observed between dark matter velocity dispersion and dark matter halo mass \cite[e.g.][]{Zahid2018}, it is unclear if  H{\sc i} emission studies can sufficiently measure properties of the larger dark matter halo associated with the host galaxy. 

Another aspect to consider is the redshift dependence of both the BTFR and the relation between rotational velocity and the dark matter velocity dispersion. This has been examined in simulations; see the review of the literature in \cite{Portinari2007}. \cite{Portinari2007} used a small sample of simulated disk galaxies ($\sim$25) to study the redshift evolution of the $B$-band TFR, and the stellar mass TFR, across redshifts of $z$~=~0, 0.7 and 1. They show in their fig.~6 that the stellar mass TFR redshift evolution is only slight, but do not comment on the baryonic TFR (that is, including the galaxy gas mass). The Millennium Simulation, plus a semi-analytic model that was used in \cite{Obreschkow2009} for a much larger sample (3~$\times$~10$^{7}$ galaxies, of various morphologies), showed the redshift evolution of the BTFR in fig. 14, panel (c), between the redshifts of $z$~=~0, 4.89, and 10.07. This study found that galaxies of equal mass rotate significantly faster at higher redshift, based on an analytical description used to generate H{\sc i} spectra. 

It is a different tale with observational studies. There do exist studies of the redshift evolution of the TFR using other measures; for instance, the $B$-band TFR is examined through optical velocity widths in \cite{Vogt2007}, and similarly the $B$, $R$, and $I$-band TFR in \cite{Lorenzo2009}. These however do not specifically consider the BTFR, nor measure it from the H{\sc i} content traced through the 21-cm transition. Observational large all-sky surveys involving direct detections of H{\sc i} through 21-cm, such as the H{\sc i} Parkes All Sky Survey \cite[HIPASS;][]{Meyer2004} and the Arecibo Legacy Fast Arecibo L-Band Feed Array (ALFALFA) survey \cite{Darling2011} have been restricted to $z$~$<$~0.1, hence limiting the ability to study the BTFR redshift evolution, among other properties.

This is set to change with upcoming SKA pathfinder telescope surveys which have improved sensitivity and hence will probe to higher redshifts. Examples include the Widefield ASKAP L-band Legacy All-sky Blind surveY \citep[WALLABY;][]{Koribalski2020} and the Deep Investigation of Neutral Gas Origins \cite[DINGO;][]{Meyer2009} survey with the Australian SKA Pathfinder telescope \cite[ASKAP;][]{Deboer2009}, and the H{\sc i} component of The MeerKAT International GHz Tiered Extragalactic Exploration (MIGHTEE) Survey \citep{Jarvis2017,Maddox2021}, MIGHTEE-H{\sc i} with the MeerKAT telescope \citep{Jonas2016}.

While many of these surveys will realistically probe out to $z\sim 0.5$, the Looking At the Distant Universe with the MeerKAT Array \cite[LADUMA;][]{Holwerda2012,Blyth2016} is set to be the deepest H{\sc i} emission survey. LADUMA will use the MeerKAT L and UHF bands to probe the evolution of gas in galaxies over cosmic time, through over 3,000 hours on a single pointing on the sky that encompasses the Extended Chandra Deep Field South (ECDFS). The deep LADUMA pointing is predicted to probe dwarf galaxies down to H{\sc i} masses of M$_{\rm HI}\sim 1\times 10^{8}$~M$_{\odot}$. This allows for direct and stacked H{\sc i} emission detections up to redshifts of $z\ga 1$.  Moreover, ECDFS has an impressive array of ancillary data in place or forthcoming, enabling correlations with a wide range of other galaxy properties. LADUMA thus will study the redshift evolution of the BTFR through H{\sc i} 21-cm observations further out than any other study. 

In preparation for LADUMA, we here make predictions for the BTFR redshift evolution using the \simba\ \citep{Dave2019} suite of cosmological hydrodynamical simulations. A local redshift study of the BTFR from galaxy rotation curves was conducted in \cite{Glowacki2020}, which found reasonable agreement with the {\em Spitzer} Photometry and Accurate Rotation Curves Survey \cite[SPARCS;][]{Lelli2016}, albeit with some interesting discrepancies in the mass dependence of the BTFR. This study included an examination of different rotational velocity methods from galaxy rotation curves, as well as dependencies on the rotation curve shape on the galaxy stellar mass and H{\sc i} gas fraction. Here we extend this work to include spectral line width measures from mock H{\sc i} data cubes across five redshift snapshots out to $z=1$, to investigate the redshift evolution of the BTFR and thus determine how viable it will be to probe this using LADUMA. Furthermore, we compare these velocity measures to the dark matter velocity dispersion and dark matter halo mass of \simba\ galaxies, to estimate how well LADUMA will be able to track these important properties. 

This paper is organised as follows. In Section~\ref{sec:sim} we describe \simba\ and outline our sample, our method of generating H{\sc i} data cubes, the observational rotational velocity measure used, and the dark matter velocity dispersion measurement. In Section~\ref{sec:results} we compare our velocity measures from rotation curves with those from spectral line profiles, and then consider the redshift evolution in the BTFR for all galaxies in our sample, and a disky galaxy subsample. Lastly, we compare these measures to the dark matter velocity dispersion, and the dark matter halo mass. We summarise our findings in Section~\ref{sec:summary}.

\section{Simulations and Analysis}\label{sec:sim}

\subsection{\simba\ }\label{sec:sample}

We employ the \simba\ simulation suite for this analysis \citep{Dave2019}. \simba\ is a cosmological hydrodynamic simulation evolved using the \gizmo\ code~\citep{Hopkins2015}, which itself is an offshoot of \gad~\citep{Springel2005}. \gizmo\ uses a meshless finite mass (MFM) hydrodynamics solver that is shown to have advantageous features over Smoothed Particle Hydrodynamics and Cartesian mesh codes, such as the ability to evolve equilibrium disks for many dynamical times without numerical fragmentation~\citep{Hopkins2015}, which is desirable for studying the rotation properties of galaxies. \simba\ also reproduces observations such as stellar growth~\citep{Dave2019}, cold gas properties~\citep{Dave2020}, and the radio galaxy population \citep{Thomas2020}.

The assumed cosmology is concordant with~\citet{Planck2016}: \(\Omega_M = 0.3\), \(\Omega_{\Lambda} = 0.7\), \(\Omega_{b} = 0.048\), \(H_0 = 68\) km s\(^{-1}\) Mpch\(^{-1}\), \(\sigma_8 = 0.82\), \(n_s = 0.97\). In this study we only consider the high mass resolution (`\simba-hires') snapshot box. \simba-hires has 8$\times$ better mass resolution and 2$\times$ better spatial resolution than the fiducial 100~Mpc~h$^{-1}$ run in \citet{Dave2019}, and so allows for more dwarf galaxies to be included -- that is, this snapshot includes lower mass galaxies compared to the fiducial 100~Mpc~h$^{-1}$ run. \simba-hires has a box size of 25~Mpc~h$^{-1}$, with 512$^3$ dark matter particles and 512$^3$ gas elements. This yields a mass resolution of \(1.2\times 10^7 M_{\odot}\) for dark matter particles and \(2.28\times 10^6 M_{\odot}\) for gas elements. Adaptive gravitational softening length is employed with a minimum Plummer equivalent softening of \(\epsilon_{\rm{min}} = 0.5h^{-1}\)c\,kpc. A description of the star formation, feedback, and accretion mechanisms employed in \simba\ is provided in \cite{Glowacki2020}, and in further detail in \cite{Dave2019}.


We initially consider snapshots at three different redshifts: $z$~=~0, 0.5 and 1. As in \cite{Glowacki2020}, in order to select galaxies with active star formation and sufficient gas and stellar particles, we impose the following limits on galaxies within these snapshots: $M_{\rm HI}$~$>$~1.25$\times$10$^{8}$~M$_{\odot}$, $M_{*}$~$>$~7.25$\times$10$^{8}$~M$_{\odot}$, and sSFR~$>$~1$\times$10$^{-11}$\,yr$^{-1}$. These limits ensure that we only consider galaxies above the stellar mass resolution limit (i.e. are resolved) for the \simba-hires snapshots, and which have sufficient H{\sc i} content to compare to observations and match the $M_{\rm HI}$ limit of LADUMA, and further have active star formation. This imposes a bias against lower stellar mass galaxies in this study, but unfortunately galaxies below this stellar mass resolution limit in \simba\ are poorly represented owing to too few ($<32$) star particles. We also visually inspect and remove galaxies undergoing obvious merger events. For our three redshift snapshots from low to high redshift, we obtain 1043, 559, and 180 galaxies. 

\begin{figure*}
\centering
\begin{minipage}{\textwidth}
\centering
\small
\begin{subfigure}[b]{0.46\textwidth}
  \includegraphics[width=1.\linewidth]{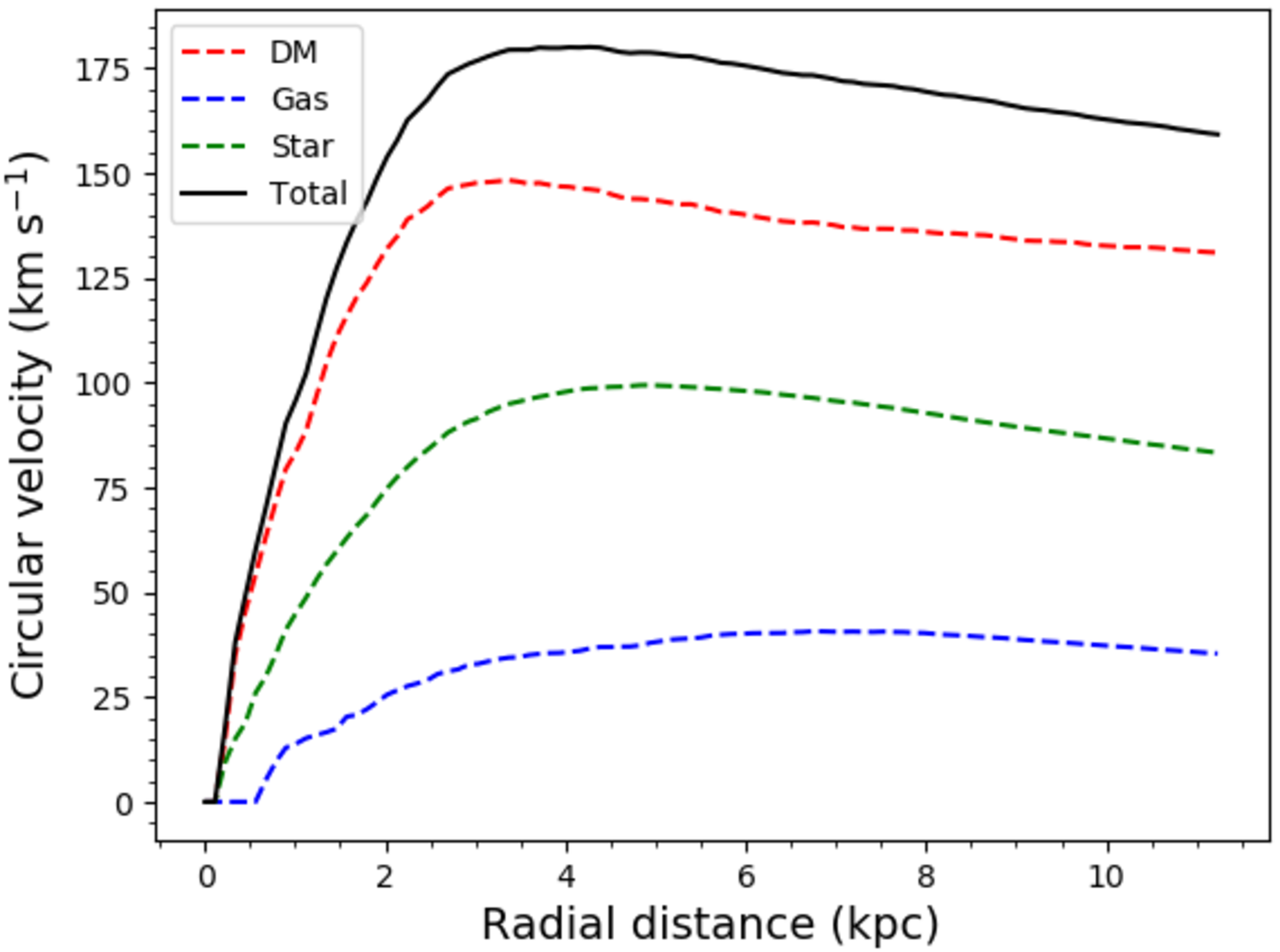}
\end{subfigure}%
\begin{subfigure}[b]{0.54\textwidth}
  \includegraphics[width=1.\linewidth]{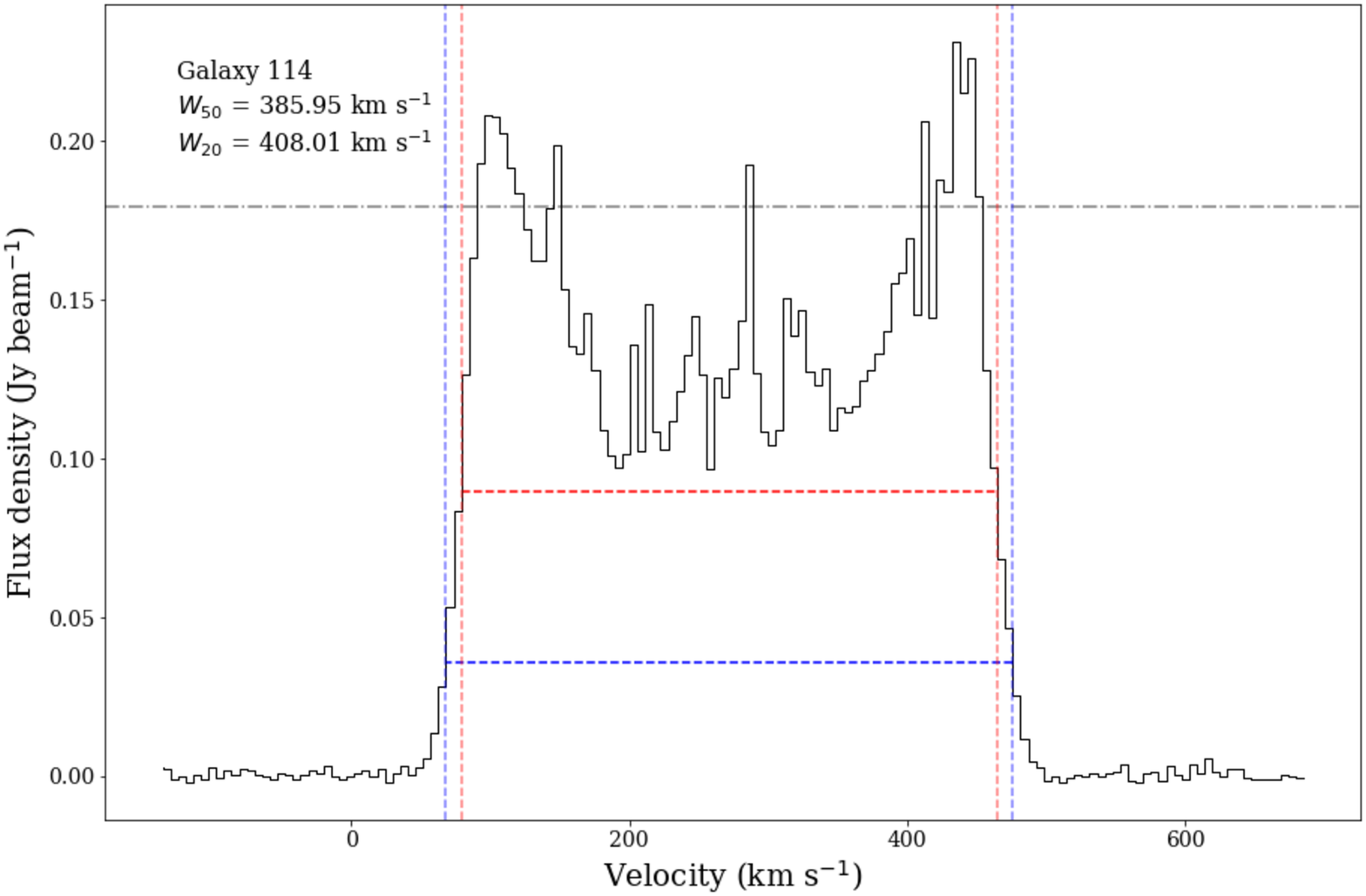}
\end{subfigure}
\begin{subfigure}[b]{0.46\textwidth}
  \includegraphics[width=1.\linewidth]{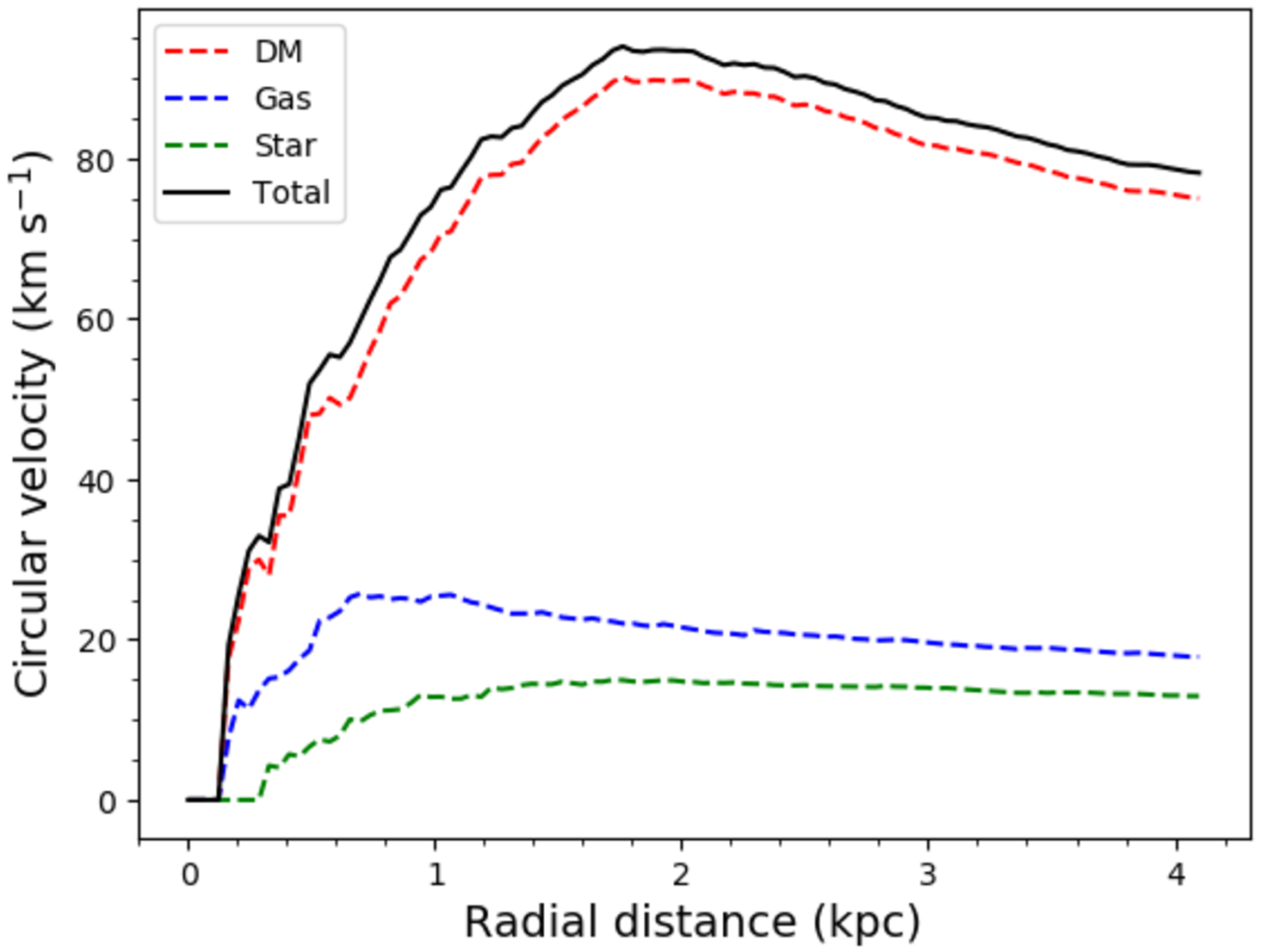}
\end{subfigure}%
\begin{subfigure}[b]{0.54\textwidth}
  \includegraphics[width=1.\linewidth]{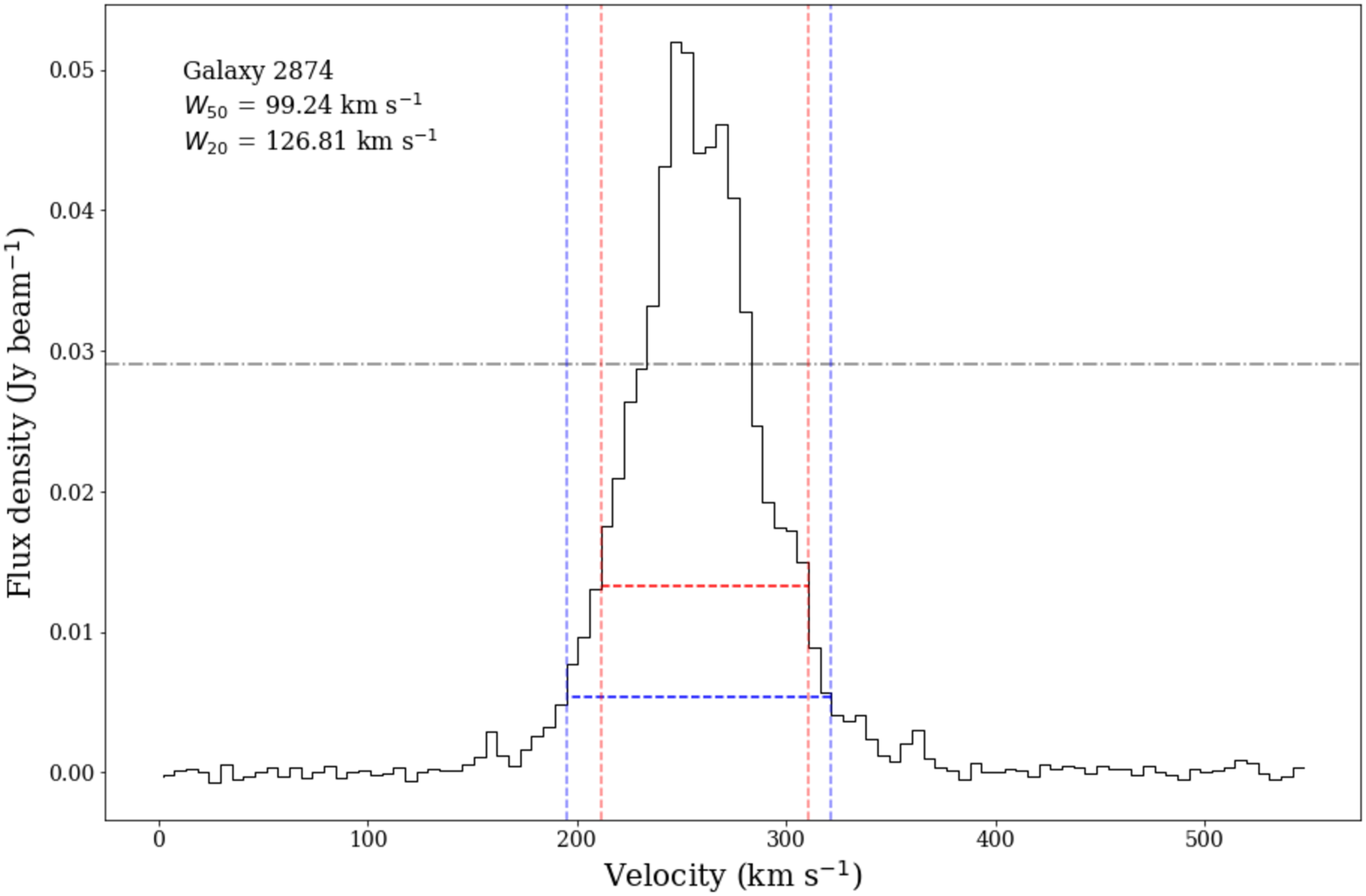}
\end{subfigure}
\caption{Rotation curves and H{\sc i} 21-cm spectra for two example  \simba\ galaxies (one massive and one dwarf galaxy, with M$_{\rm HI}$~$\sim$~4$\times$10$^{9}$ and $\sim$~2$\times$10$^{8}$~M$_{\odot}$ respectively) from the high resolution snapshot at $z$~=~0. The rotation curves are computed as described in \citet{Glowacki2020}. H{\sc i} spectra are obtained from a H{\sc i} cube created through \martini\ with the spectral resolution of MeerKAT at 1420 MHz (5.51~km\,s$^{-1}$), and convolved with a standard radio beam and noise. The $W_{\rm 50}$ and $W_{\rm 20}$ values - the width at 50\% and 20\% of the 90th percentile of the flux - are overplotted. The 90th percentile flux value is indicated by the grey dot-dashed line.}
\label{fig:hispectra}
\end{minipage}
\end{figure*}

Our sample sizes are roughly comparable to the expected numbers of galaxies to be detected in LADUMA: below redshifts of $z$~$<$~0.4, 630 to 1,300 detections are predicted (dependent on the effect of radio frequency interference (RFI)-afflicted frequencies), between redshifts 0.4--0.56 $\sim$1,200 sources, and between redshifts 0.56--1.4 around $\sim$700 galaxies in total (based on Oxford S$^{3}$ simulations; \citealt{Obreschkow2009b}). The field of view of LADUMA of 2~deg$^{2}$ at $z~\sim~0.58$ \citep{Blyth2016} corresponds to 48~Mpc, which is comparable to \simba-hires' box size of 25 Mpc h$^{-1}$ (roughly 37~Mpc at each redshift) -- we note that as redshift increases, the LADUMA field of view increases, owing to observing \HI\ at lower frequencies. By not using the larger volume snapshot galaxies, we also minimise the number of massive ($M_{*}$~$>$10$^{10}$~M$_{\odot}$) galaxies, which we do not expect to see large numbers of in LADUMA compared to shallower H{\sc i} surveys covering a larger area, such as MIGHTEE-HI \citep{Jarvis2017}. LADUMA is also predicted to probe dwarf galaxies down to H{\sc i} masses of M$_{\rm HI}\sim 1\times 10^{8}$~M$_{\odot}$ at low redshifts, making \simba-hires a good sample to use for predictions for LADUMA.

From these \simba\ galaxies, we also consider a further subsample during the analysis based on galaxy morphology. While LADUMA will detect galaxies in all morphologies in their H{\sc i} emission, BTFR studies have focused on the disk galaxy population, given elliptical galaxies, which are kinematically dominated by velocity dispersion rather than rotational velocity, do not observe the same tight relation. To distinguish between disky galaxies and otherwise, we employ the $\kappa_{\rm rot}$ measure, the fraction of kinetic energy invested in ordered rotation, as described in \cite{Sales2012}:

\begin{equation}
    \kappa_{\rm rot} = \frac{K_{\rm rot}}{K} = \frac{1}{K} \sum \frac{1}{2} m \left(\frac{j_{\rm z}}{r}\right)^{2},
\end{equation}
where $j_{\rm z}$ is the specific angular momentum perpendicular to the disc. In that work, they characterise simulated galaxies with $\kappa_{\rm rot}$~$<$~0.5 as spheroid-dominated, and $\kappa_{\rm rot}$~$>$~0.7 as disk-dominated. Therefore, we employ the same $\kappa_{\rm rot}$~$>$~0.7 limit to select our disky galaxies for each snapshot. Roughly 15--20\% of our \simba\ galaxies in each redshift sample qualify as disk-dominated. We have verified that this matches our visual inspection of galaxies, and given this and the results of \cite{Sales2012}, we expect results will agree well with disky galaxies in LADUMA. We note LADUMA will rely on multi-wavelength follow-up to identify disky galaxies, and will not by itself measure $\kappa_{\rm rot}$.

In order to better examine the BTFR for disky galaxies, we also consider disky galaxies from the snapshots at $z$~=~0.25 and 0.75. These samples contain 136 and 53 galaxies respectively. These galaxies underwent the same selection criteria and process as the disky $z$~=~0, 0.5 and 1 samples.

\subsection{HI spectra}

We generate H{\sc i} cubes via \martini\footnote{\url{https://github.com/kyleaoman/martini}}. \martini\ is a package for creating synthetic resolved HI line observations -- aka data cubes -- of smoothed particle hydrodynamical simulations of galaxies. It allows for realistic mock observations in that one can set the spectral resolution, cube size and galaxy inclination relative to the observer, and convolve the beam with a radio beam and incorporate noise into the output data cube. \martini\ takes the input snapshot file and accompanying \caesar\footnote{\tt https://caesar.readthedocs.io/en/latest/} catalog, a particle-based extension to {\sc yt}\footnote{\tt https://yt-project.org/}, which contains the galaxy and host halo properties including H{\sc i} fraction values. \martini\ also accepts user-defined measures of the galaxy distance, rotation, and radio observation properties (spatial and spectral cube resolution, noise, and radio beam dimensions) as inputs. We note that the H{\sc i} fraction in each gas particle is computed in \simba, accounting for self-shielding on the fly based on the prescription in \cite{Rahmati2013}, and includes photoionisation from a spatially uniform ionising background given by \cite{Haardt2012}.

We opt to use properties that match the 32K mode of the MeerKAT telescope, to be employed for the upcoming LADUMA and MIGHTEE-HI surveys. That is, we use a spectral resolution of 5.51~km\,s$^{-1}$, a common radio beam of $\sim$10 arcseconds (BMAJ and BMIN, the major and minor axis of the radio beam, of 11.2 and 9.8~arcseconds, as per early L-band MeerKAT observations of LADUMA), and noise values of 5$\times$10$^{-8}$~Jy\,arcsec$^{-2}$ before convolution. All galaxies were inclined such to be edge on to the observer, so that no inclination correction is required to the rotational velocity measures. 

We extract an H{\sc i} spectrum from each data cube through spectral-cube\footnote{\url{https://spectral-cube.readthedocs.io/en/latest/}} (Python module) , which has also been employed in early science and data verification analysis of LADUMA. Fig.~\ref{fig:hispectra} show two spectra for a large and a small galaxy (M$_{\rm HI}$~$\sim$~4$\times$10$^{9}$ and $\sim$~2$\times$10$^{8}$~M$_{\odot}$) in the right panels. Alongside each spectrum in the left panels are each galaxy's corresponding rotation curve as constructed in \cite{Glowacki2020}, based on the enclosed mass at a given radius. The larger galaxy shows a double-horned profile typical of disk galaxy rotation curves.  The smaller galaxy is also identified as rotationally supported, but has a less well defined disk and shows a more Gaussian-like profile.

\subsection{Rotational velocity measures}

\begin{figure*}
\centering
\begin{subfigure}[b]{0.89\textwidth}
  \includegraphics[width=1.\linewidth]{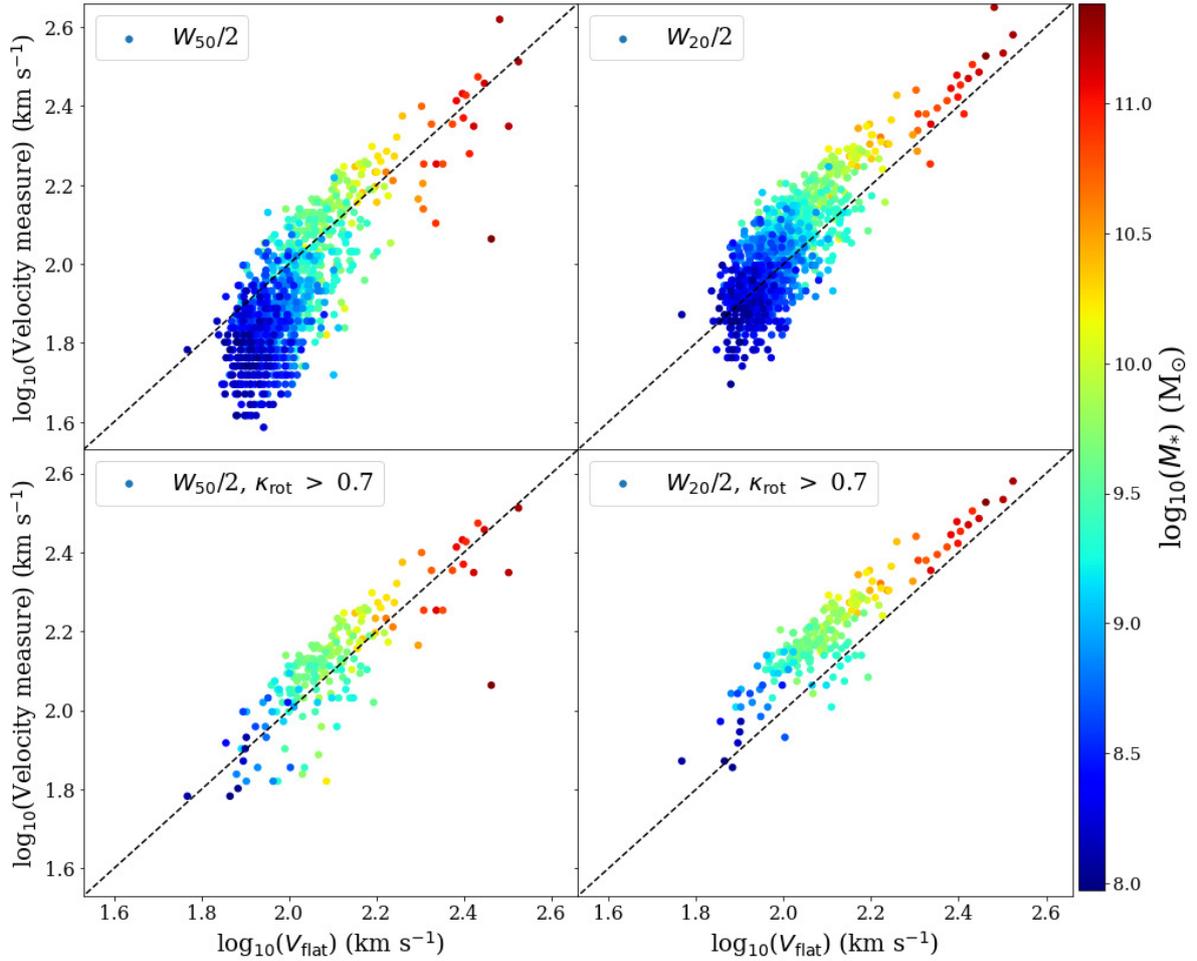}
\end{subfigure}
\caption{$W_{\rm 50}$ (left) and $W_{\rm 20}$ (right) versus $V_{\rm flat}$, coloured by galaxy stellar mass, for \simba\ galaxies. The top panels are for the full sample, and bottom panels disky galaxies. The 1:1 relation is given in the black dashed lines.}
\label{fig:w50w20_vs_vflat}
\end{figure*}

We consider three rotational velocity measures from the rotation curves and spectra that are used in observational studies:

\begin{itemize}
    \item $V_{\rm flat}$ - the average circular velocity along the flat part of the rotation curve, defined in \cite{Lelli2016} as the average of outermost points of a rotation curve with relative cumulative differences smaller than 5\% in rotational velocity. We choose $V_{\rm flat}$ as this measure was found to give the tightest (least amount of scatter) BTFR than other velocity measures, both for the SPARC observational sample \citep{Lelli2019} and \simba\ galaxies \citep{Glowacki2020}. Furthermore, we find a better agreement between the SPARC and \simba\ samples for this definition, which is less sensitive than e.g. $V_{\rm max}$ to an excess of dark matter content in the cores of SIMBA galaxies. To determine the maximum radius for the rotation curve, we use the \HI\ mass-size relation to calculate the expected radius of each galaxy, at an H{\sc i} mass surface density level of 1~M$_{\odot}$\,pc$^{-2}$. There exists a tight relation between the \HI\ mass and the extent of the \HI\ disk for spiral galaxies in observational work \citep{Broeils1997,Verheijen2001,Swaters2002,Noordermeer2005,Wang2016}. We apply the relation given in \cite{Wang2016} to the \HI\ masses we have within \simba\ for each galaxy to derive a \HI\ radius.
    \item $W_{\rm 50}$/(2sin($i$)) - the width at 50\% of a flux value, divided by two as per observational studies \cite[e.g.][]{Ponomareva2017,Lelli2019}. We note that in \martini\ all cubes are rotated so that the inclination angle $i$~=~90\,$\degree$ for each galaxy, and so sin($i)$ = 1. Therefore, we henceforth refer to this measure as $W_{\rm 50}$/2. We opt to use the 90th percentile of the highest flux measurement as opposed to the peak flux, as this results in a lower spread of values, but the main conclusions do not change based on either approach.
    \item $W_{\rm 20}$/2 - as with $W_{\rm 50}$/2, but half the width at 20\% of the 90th percentile of the peak flux.
\end{itemize}

Fig.~\ref{fig:hispectra} (right panels) shows the computation of $W_{\rm 50}$ (vertical red dashed line) and $W_{\rm 20}$ (vertical blue dashed line) for our two example spectra.

We also obtain, as defined in the \caesar\ files of SIMBA, the dark matter halo velocity dispersion (henceforth $\sigma_{\rm DM}$), to compare the above velocity measures with directly, and see how these observational measures serve in place of $\sigma_{\rm DM}$ in the BTFR. \caesar\ uses a friends-of-friends algorithm to identify halos, with a linking length of 0.2 times the mean inter-particle spacing. We consider both dark matter particles in the host galaxy's dark matter halo, as well as only dark matter particles within the radius from the galaxy's centre of mass used as the cut-off distance for the galaxy's rotation curve (ergo, determined by the H{\sc i} mass-size relation, using the $M_{\rm HI}$ from \caesar\ and the relation found in \citealt{Wang2016}, as done for the rotation curves in \citealt{Glowacki2020}), which it typically much smaller than the virial radius. Unless mentioned otherwise, we typically consider the $\sigma_{\rm DM}$ from the whole galaxy dark matter halo.

\section{Results}\label{sec:results}


\subsection{Rotational velocity comparisons}

Before we consider the redshift evolution in the BTFR, we want to compare our spectral line width measures, $W_{\rm 50}$/2 and $W_{\rm 20}$/2, with the $V_{\rm flat}$ measure from rotation curves. We do not expect a perfect correlation between these different measures. $W_{\rm 50}$ and $W_{\rm 20}$ measure different parts of the H{\sc i} spectrum and so, as the spectrum shape changes from the classic double peaked shape for massive galaxies to single peaked shapes (see examples in Fig.~\ref{fig:hispectra}), the difference between $W_{\rm 50}$ and $W_{\rm 20}$ also evolves. However, any deviation identified may inform us on where other relations can break down for our simulated galaxy sample.

In Fig.~\ref{fig:w50w20_vs_vflat} we plot the spectral width measurements against $V_{\rm flat}$ for the redshift $z$~=~0 snapshot. $W_{\rm 50}$/2 occupies the left-side panels, and $W_{\rm 20}$/2 the right. The top panels are for the full sample, and bottom panels the disky subsample. Points are coloured by the stellar mass, and the 1:1 relation is given by the black dashed lines. 

Considering the full sample first, $W_{\rm 50}$/2 follows the 1:1 relation at higher masses (M$_{*}$~$>$~10$^{9.5}$~M$_{\odot}$), but at lower masses in the full sample (or alternatively below rotational velocities of $\sim$100~km\,s$^{-1}$) $W_{\rm 50}$/2 gives lower rotational velocities than $V_{\rm flat}$. While this effect is also present for $W_{\rm 20}$/2, it is far less pronounced. It is possible that it is because the $W_{\rm 20}$/2 measure, made at a lower section of the spectrum, is less sensitive to changes in the spectrum shape than width measures made closer to the peak flux, as $W_{\rm 50}$/2 is. We also find a lower scatter from $W_{\rm 20}$/2 than for $W_{\rm 50}$/2, which holds true throughout this study when either are used to compare different relations. 

\begin{table*}
    \centering
    \begin{tabular}{lccccc}
    \hline 
Velocity definition & Redshift & Sample size & $m$ & $b$ & $\sigma_{\perp}$\\
\hline
Full sample\\
\hline
$V_{\rm flat}$ & 0 & 1043 & 4.38$\pm$0.07 & 0.56$\pm$0.13 & 0.070$\pm$0.004\\
$V_{\rm flat}$ & 0.5 & 559 & 3.67$\pm$0.14 & 2.34$\pm$0.28 & 0.078$\pm$0.011\\
$V_{\rm flat}$ & 1 & 180 & 3.64$\pm$0.12 & 2.60$\pm$0.25 & 0.075$\pm$0.014\\
\hline
$\sigma_{\rm DM}$ & 0 & 1043 & 3.54$\pm$0.09 & 2.55$\pm$0.17 & 0.073$\pm$0.007\\
$\sigma_{\rm DM}$ & 0.5 & 559 & 2.85$\pm$0.08 & 3.96$\pm$0.16 & 0.101$\pm$0.014\\
$\sigma_{\rm DM}$ & 1 & 180 & 3.22$\pm$0.08 & 3.06$\pm$0.16 & 0.066$\pm$0.008\\
\hline
$W_{\rm 50}$/2 & 0 & 1043 & 2.60$\pm$0.04 & 4.32$\pm$0.08 & 0.115$\pm$0.008\\
$W_{\rm 50}$/2 & 0.5 & 559 & 2.27$\pm$0.08 & 5.17$\pm$0.15 & 0.143$\pm$0.019\\
$W_{\rm 50}$/2 & 1 & 180 & 2.20$\pm$0.12 & 5.41$\pm$0.25 & 0.159$\pm$0.027\\
\hline
$W_{\rm 20}$/2 & 0 & 1043 & 3.50$\pm$0.04 & 2.22$\pm$0.08 & 0.068$\pm$0.004\\
$W_{\rm 20}$/2 & 0.5 & 559 & 2.64$\pm$0.12 & 3.97$\pm$0.24 & 0.090$\pm$0.019\\
$W_{\rm 20}$/2 & 1 & 180 & 2.97$\pm$0.08 & 3.40$\pm$0.17 & 0.092$\pm$0.011\\
\hline
Disky galaxy subsample ($\kappa_{\rm rot}$~$>$~0.7)\\
\hline
$V_{\rm flat}$ & 0 & 204 & 3.62$\pm$0.12 & 2.42$\pm$0.25 & 0.086$\pm$0.013\\
$V_{\rm flat}$ & 0.25 & 136 & 3.33$\pm$0.09 & 3.17$\pm$0.18 & 0.094$\pm$0.010\\
$V_{\rm flat}$ & 0.5 & 86 & 3.91$\pm$0.22 & 1.93$\pm$0.46 & 0.073$\pm$0.017\\
$V_{\rm flat}$ & 0.75 & 51 & 3.78$\pm$0.24 & 2.29$\pm$0.51 & 0.056$\pm$0.004\\
$V_{\rm flat}$ & 1 & 33 & 4.43$\pm$0.38 & 0.94$\pm$0.81 & 0.047$\pm$0.025\\
\hline
$\sigma_{\rm DM}$ & 0 & 204 & 3.27$\pm$0.06 & 3.27$\pm$0.12 & 0.078$\pm$0.007\\
$\sigma_{\rm DM}$ & 0.25 & 136 & 2.78$\pm$0.09 & 4.27$\pm$0.19 & 0.104$\pm$0.015\\
$\sigma_{\rm DM}$ & 0.5 & 86 & 2.86$\pm$0.12 & 3.95$\pm$0.26 & 0.120$\pm$0.027\\
$\sigma_{\rm DM}$ & 0.75 & 51 & 3.01$\pm$0.14 & 3.56$\pm$0.32 & 0.056$\pm$0.007\\
$\sigma_{\rm DM}$ & 1 & 33 & 3.27$\pm$0.18 & 2.92$\pm$0.40 & 0.061$\pm$0.016\\
\hline
$W_{\rm 50}$/2 & 0 & 204 & 3.02$\pm$0.15 & 3.52$\pm$0.31 & 0.111$\pm$0.028\\
$W_{\rm 50}$/2 & 0.25 & 136 & 2.67$\pm$0.31 & 4.21$\pm$0.48 & 0.148$\pm$0.056\\
$W_{\rm 50}$/2 & 0.5 & 86 & 2.36$\pm$0.21 & 4.88$\pm$0.47 & 0.162$\pm$0.074\\
$W_{\rm 50}$/2 & 0.75 & 51 & 4.83$\pm$0.48 & -0.62$\pm$1.10 & 0.093$\pm$0.060\\
$W_{\rm 50}$/2 & 1 & 33 & 5.02$\pm$0.59 & -1.21$\pm$1.36 & 0.065$\pm$0.086\\
\hline
$W_{\rm 20}$/2 & 0 & 204 & 4.06$\pm$0.10 & 1.07$\pm$0.23 & 0.057$\pm$0.007\\
$W_{\rm 20}$/2 & 0.25 & 136 & 3.92$\pm$0.17 & 1.32$\pm$0.32 & 0.070$\pm$0.017\\
$W_{\rm 20}$/2 & 0.5 & 86 & 4.02$\pm$0.18 & 0.99$\pm$0.39 & 0.073$\pm$0.011\\
$W_{\rm 20}$/2 & 0.75 & 51 & 4.28$\pm$0.25 & 0.33$\pm$0.42 & 0.048$\pm$0.017\\
$W_{\rm 20}$/2 & 1 & 33 & 5.09$\pm$0.27 & -1.65$\pm$0.59 & 0.051$\pm$0.018\\
\hline
    \end{tabular}
    \caption{The best fit BTFRs for different rotational velocity definitions: $V_{\rm flat}$ from rotation curves, $W_{\rm 50}$/2 and $W_{\rm 20}$/2 from spectral line widths, and the dark matter velocity dispersion ($\sigma_{\rm DM}$), for different redshift snapshots. We give the slope ($m$), y-intercept ($b$), and the orthogonal intrinsic scatter ($\sigma_{\perp}$) in dex for log$_{\rm{10}}$(M$_{\rm{bar}}$) = m\,log$_{\rm{10}}$($V$) + b. The standard deviation from 50 sets of bootstrapping on each sample is given as an error for each parameter. We compare between the full sample (which requires a minimum H{\sc i} mass, stellar mass, and sSFR), and a disky subsample which additionally requires $\kappa_{\rm rot}$~$>$~0.7, at different redshifts.}
    \label{tab:btfr_fits}
\end{table*}
    

\begin{figure*}
\centering
\begin{subfigure}[b]{0.76\textwidth}
  \includegraphics[width=1.\linewidth]{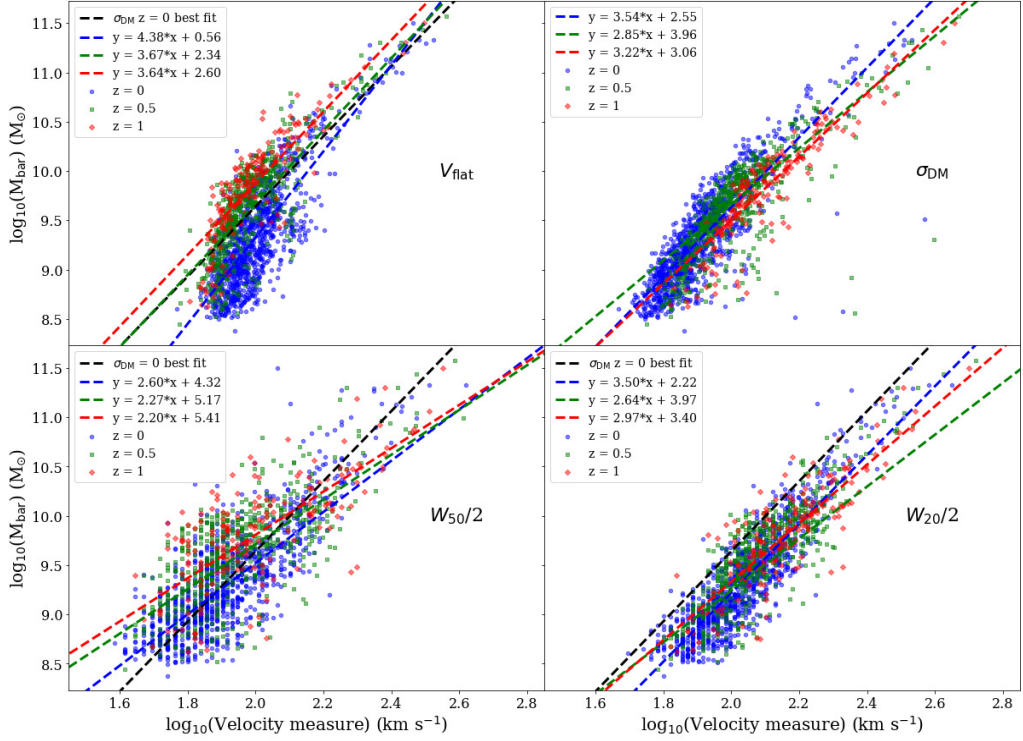}
\end{subfigure}
\caption{The redshift evolution of the BTFR for $V_{\rm flat}$, $\sigma_{\rm DM}$, $W_{\rm 50}$/2 and $W_{\rm 20}$/2, for all galaxies selected from each snapshot (not specifically just disky galaxies). A simple linear fit is given in dashed lines; we note the relation steepens for $V_{\rm flat}$ at lower baryonic masses. Visually the distributions are distinct for the top two panels, but harder to separate when using spectral line widths due to contamination of non-disk galaxies. The $W_{\rm 50}$/2 measure has a significant scatter.}
\label{fig:btfr_redshift}
\end{figure*}

\begin{figure*}
\centering
\begin{subfigure}[b]{0.76\textwidth}
  \includegraphics[width=1.\linewidth]{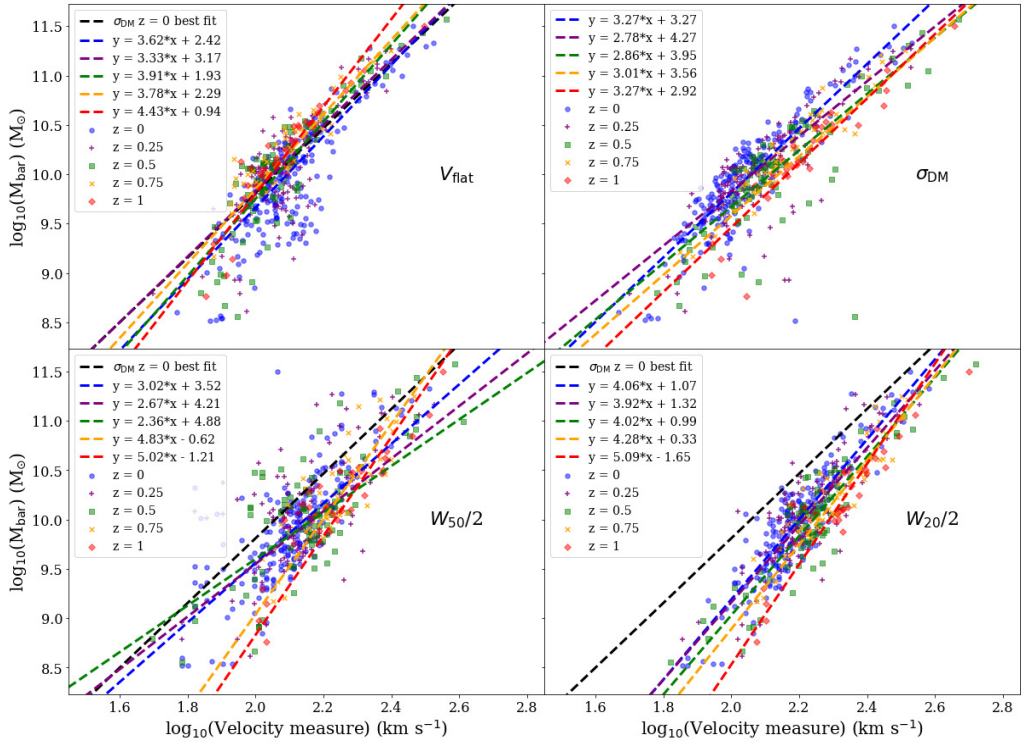}
\end{subfigure}
\caption{As in Fig.~\ref{fig:btfr_redshift}, but with a cut for disky ($\kappa_{\rm rot}$~$>$~0.7) galaxies, and including galaxies at $z$~=~0.25 and 0.75. The redshift evolution is more pronounced in the $W_{\rm 20}$/2 measure for these disky galaxies. The scatter in $W_{\rm 50}$/2 is too large for convincing redshift evolution to be seen. The improvement in these relations compared to Fig.~\ref{fig:btfr_redshift} showcases the importance for galaxy morphological identification in LADUMA.}
\label{fig:btfr_redshift_krot}
\end{figure*}

\begin{figure}
\centering
\begin{subfigure}[b]{0.49\textwidth}
  \includegraphics[width=0.99\linewidth]{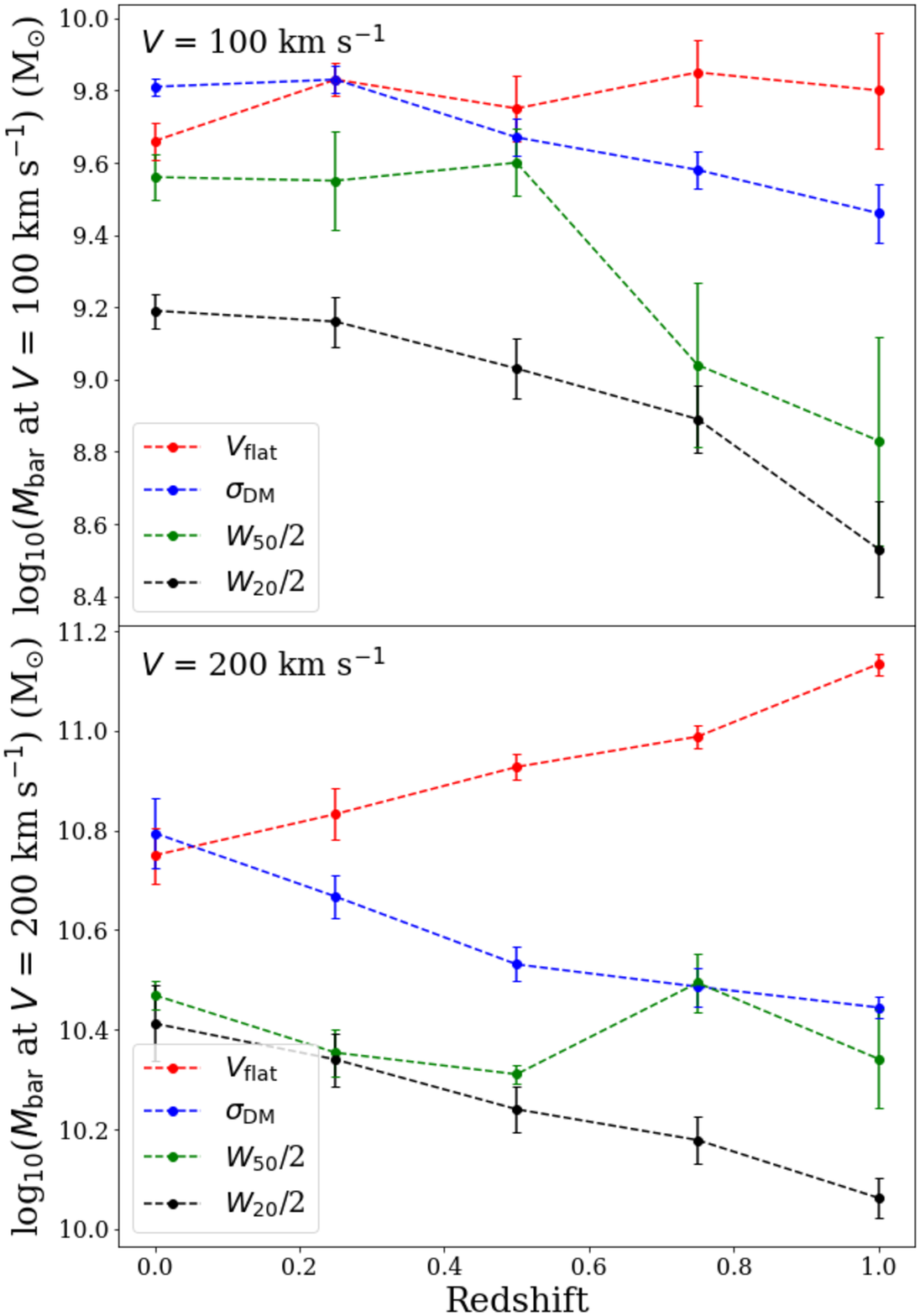}
\end{subfigure}%
\caption{The baryonic mass $M_{\rm bar}$ for the best-fit relations for disky galaxies (Fig.~\ref{fig:btfr_redshift_krot}) versus redshift, at rotational velocities of 100~km\,s$^{-1}$ (top panel) and 200~km\,s$^{-1}$ (bottom panel). At both velocities $\sigma_{\rm DM}$ and $W_{\rm 20}$ decrease in $M_{\rm bar}$ with redshift, while $V_{\rm flat}$ increases.}
\label{fig:bmass_fixedvel}
\end{figure}

\begin{figure*}
\centering
\begin{subfigure}[b]{0.86\textwidth}
  \includegraphics[width=1.\linewidth]{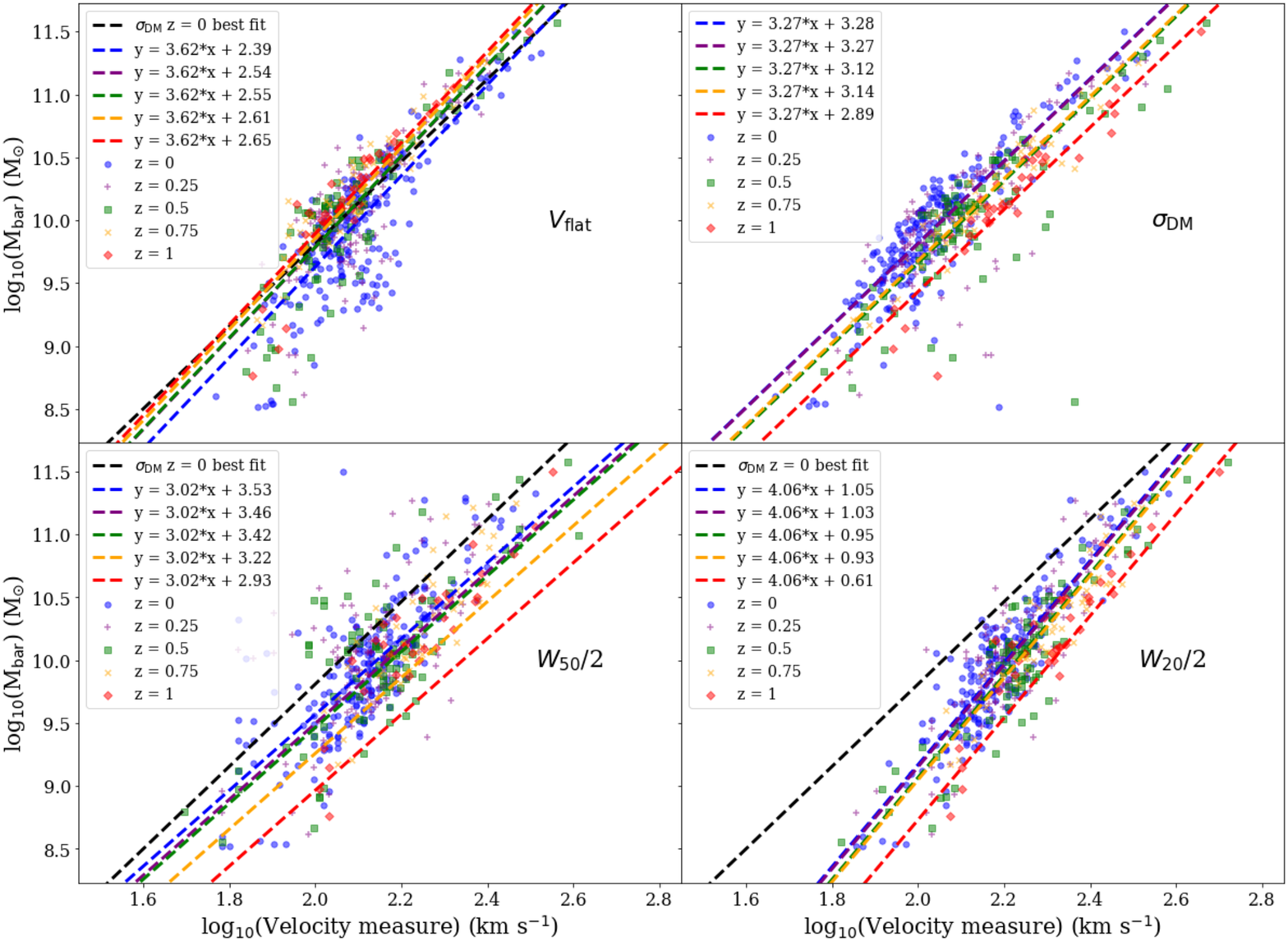}
\end{subfigure}
\caption{As in Fig.~\ref{fig:btfr_redshift_krot}, but where the slope for each best-fit relation is fixed to the $z = 0$ case for all redshift samples. The best-fit y-intercept slightly decreases with redshift for $\sigma_{\rm DM}$ and the spectral line width measures, while it increases in the case of $V_{\rm flat}$.}
\label{fig:btfr_redshift_fixedslope}
\end{figure*}
 
We rule out that the higher $V_{\rm flat}$ measures at low masses in the full sample is driven by the rotation curve shape; unlike in observational studies, the rotation curves for our low-mass SIMBA galaxies still exhibit a peak and flattening in their shape, rather than continually rising, which is an issue with \simba\ that was noted in \citet{Glowacki2020}. This may be simply because we are not affected in sensitivity with simulated galaxies to reach the H{\sc i} mass-size relation distance used to determine the extent of the rotation curve, which may be an issue for observational studies for lower-mass galaxies. 

To verify this, we fitted the Polyex parametric model \citep{Giovanelli2002} to all rotation curves in our sample here (see fig.~4 of \cite{Glowacki2020}). The Polyex parametric model $V_{\rm{polyex}}$ is defined as

\begin{equation}
V(r) = V_{0}(1 - e^{-r/r_{\rm{pe}}})(1 - \alpha r/r_{\rm{pe}}),    
\end{equation}
where $V_{\rm{0}}$ regulates the overall amplitude of the rotation curve, $r_{\rm{pe}}$ is the scale length for the inner steep rise of the rotation curve, and yields a scale length for the inner steep rise, and $\alpha$ sets the slope of the slowly varying outer part. No trend was found with stellar mass between the $r_{\rm pe}$ parameter of Polyex with the radius at which the maximum velocity of the rotation curve occurs, normalised by the extent of the rotation curve's radius. Therefore, there is no evidence that the higher peak in \simba\ galaxy rotation curves, which corresponds to the radius at which the peak $V_{\rm 0}$ occurs in Polyex, is caused by the rotation curve shape itself. Furthermore, the same downturn between spectral line width and $V_{\rm flat}$ is seen for other velocity measures from the rotation curve, e.g. $V_{\rm max}$.

The disky sample (bottom panels) better follows the 1:1 relation and lacks the downturn at low $V_{\rm flat}$ values, particularly for the smaller galaxies. This indicates that the linewidths fail to trace $V_{\rm flat}$ in low-mass, dispersion-dominated galaxies, which is perhaps unsurprising because the majority of those are not disky. Instead, low-mass \simba\ galaxies are more supported by dispersion than by rotation, as seen in elliptical and irregular galaxies. Therefore, for lower mass galaxies in the full sample we can potentially expect to see some deviation from any trend found for the BTFR or with $\sigma_{\rm DM}$.

Fitting a line with unity slope to the disky sample, we get an amplitude offset of about 0.05~dex for the $W_{\rm 50}$ measure and 0.2~dex for $W_{\rm 20}$ relative to $V_{\rm flat}$. This shows that $W_{\rm 50}$ nicely traces $V_{\rm flat}$ as expected, while $W_{\rm 20}$ does also albeit with a small offset. Nonetheless, $W_{\rm 20}$ provides a tighter relation versus $V_{\rm flat}$ (orthogonal intrinsic scatter of $\sigma_\perp$~=~0.039 for $W_{\rm 20}$, versus $\sigma_\perp$~=~0.057 for $W_{\rm 50}$; see Section~\ref{sec:btfr_allfit} for further details on $\sigma_\perp$ measurements). Thus both these measures are useful for tracing the rotation velocities of galaxies in lieu of resolved data.


Overall for \simba\ galaxies we see good agreement between our rotational velocity measures from spectral line widths and rotation curves, albeit with a fixed offset. Next, we consider how each of these measures, alongside $\sigma_{\rm DM}$, compare with the BTFR at different redshifts.

\subsection{Redshift evolution of the BTFR}

Armed with our various rotation measures, we can now examine the redshift evolution of the BTFR.  We now also include $\sigma_{\rm DM}$ as a rotation measure, in order to understand how well this quantity is traced by HI-based galaxy rotation measures. All plots on the BTFR use $M_{\rm bar}$, the baryonic mass of each galaxy (i.e. the cool, dense gas plus stellar mass). By redshift evolution, we refer to both the slope, and $y$-intercept (or normalisation) of the BTFR.

\subsubsection{HI-rich star-forming galaxies}\label{sec:btfr_allfit}

We begin in Fig.~\ref{fig:btfr_redshift}, by considering the full sample selected to have significant \HI, as described in Section~\ref{sec:sample}. While this sample includes some non-disky galaxies, this is the sample that represents what will be directly identified from LADUMA. Disky galaxies from LADUMA will need to be identified with ancillary data, and so we wish to first consider the larger \simba\ sample in order to highlight the importance for LADUMA to identify spiral galaxies. We plot the BTFR for each rotational velocity measure and $\sigma_{\rm DM}$ in each panel, on a log-log plot. Redshift $z$~=~0 results are given in blue, $z$~=~0.5 in green, and $z$~=~1 in red.  Also shown are the best-fit linear relations, in matched colours. 

Table~\ref{tab:btfr_fits} gives the values and 1$\sigma$ errors for the best fitting linear values for slope $m$, y-intercept $b$, and their orthogonal intrinsic scatter $\sigma_\perp$. The measure of scatter is quantified as in \cite{Lelli2019} and \cite{Glowacki2020} through the standard affine-invariant ensemble sampler in {\sc emcee} \citep{Foreman-Mackey2013}, which includes inverse variance weighting. The best fit relation from the $\sigma_{\rm DM}$ measure at $z$~=~0 is given in each panel for comparison, as the black dashed line. We note that at times in our efforts to simulate the LADUMA survey through this sample, that the sample size is limited, especially for the disky subsample (see second half of Table~\ref{tab:btfr_fits}). Therefore, in estimating the errors, we both use 100 random walkers through the {\sc emcee} fitting routine, and repeat on fifty sets of bootstrapping (random sampling with replacement) to obtain realistic error estimates. Errors quoted for the slope, y-intercept, and scatter are taken from the standard deviation of our bootstrapping results. Our measured median values for these parameters do not change significantly between the bootstrapping approach and a single {\sc emcee} analysis of the original set. 

In regards to scatter, for all redshifts we see $W_{\rm 50}$/2 has the largest spread, which is also evident visually, particularly at the low baryonic mass or velocity end. This likely also contributes to the significantly flatter BTFR slopes we see for this measure. The results for $W_{\rm 50}$/2 is attributed to low velocity values, which may be due to the inclusion of non-disky galaxies. Another effect could be the discrete spectral resolution of the H{\sc i} data cube. As shown in Fig.~\ref{fig:hispectra}, $W_{\rm 50}$/2 will be smaller than $W_{\rm 20}$/2, and so the effect is lessened for $W_{\rm 20}$/2 for galaxies with low baryonic mass.

The other two observational measures, and the non-observational method $\sigma_{\rm DM}$, have similar values of scatter as each other. This suggests that for better results in LADUMA, velocity measures that traces a wider width from the H{\sc i} spectral line profile is recommended at higher redshift, and for local galaxies measures from rotation curves. We note that in \cite{Lelli2019} they use $W_{\rm M50}$, the line width at 50\% of the mean flux density, rather than the peak, ergo a value comparable to $W_{\rm 20}$, and found similar measures of scatter for $W_{\rm 20}$/2 and $W_{\rm M50}$/2 for SPARC galaxies.

As mentioned, this full sample does not contain only disky galaxies, but also includes dispersion-dominated galaxies. We thus do not focus on analysis for this sample, beyond making the observation that the large scatter prevents a clear redshift evolution of the BTFR to be apparent through the spectral line width measures. As galaxies at higher redshift will not be spatially resolved by MeerKAT, measures from rotation curves such as $V_{\rm flat}$ will not prove a practical method. This means spectral line widths must be relied upon in a redshift evolution study. As shown here these measures are not ideal for tracing $V_{\rm flat}$ based on the full sample from \simba\ - therefore, LADUMA must be able to distinguish between disky and elliptical morphologies. Happily, this is not the end of the tale.

\subsubsection{Disky galaxies}\label{sec:btfr_disky}

The sample considered earlier were merely star-forming galaxies with sufficient H{\sc i} gas, not just disky galaxies, which is what the BTFR focuses on. While LADUMA will detect H{\sc i} emission in galaxies of all morphologies, it will be able to discern between host galaxy morphology only due to the wealth of multi-wavelength data available in the ECDF-S \citep{Holwerda2012}. Hence in Fig.~\ref{fig:btfr_redshift_krot} we consider the subsample of \simba\ galaxies with disky morphologies, using the $\kappa_{\rm rot}$~$>$~0.7 limit. Their best fits are also included in Table~\ref{tab:btfr_fits}. We note that the same trend in scatter is seen for the disky subsample as in the full sample. We include the disky galaxies in the $z$~=~0.25 and 0.75 snapshots in this analysis, so to better examine the trend within redshifts 0 to 1.

First, we consider $\sigma_{\rm DM}$, which the BTFR aims to trace through observational methods. The $z~=~0.5$ galaxies have a flatter slope than for redshift $z = 0$, while the slope for $z~=~1$ is within 1$\sigma$ of the best-fit slope of the $z~=~0$ sample. The y-intercept meanwhile grows between $z~=~0\rightarrow 0.5$, but drops again to $z~=~1$. Visually, we see that the higher redshift galaxies tend to occupy the lower right side of the plot. To better quantify this, in Fig.~\ref{fig:bmass_fixedvel} we give the baryonic mass for the best-fit relation at fixed velocities of 100 and 200~km\,s$^{-1}$. For $\sigma_{\rm DM}$ (black points), the baryonic mass decreases with increasing redshift (or alternatively, at fixed baryonic mass, galaxies rotate faster at higher redshift).

Curiously, we see that the visual redshift evolution trends are reversed for $V_{\rm flat}$; when using $\sigma_{\rm DM}$, the higher redshift galaxies have larger dark matter velocity dispersions than at lower redshift, while with $V_{\rm flat}$ it is the lower redshift galaxies with higher rotational velocities. Quantitatively, $V_{\rm flat}$ has an increase in baryonic mass at fixed velocity with redshift, albeit this is weak between $z$~=~0 to 1 at lower fixed velocities; the trend is more convincing ($>3\sigma$ significance) at $V$~=~200~km\,s$^{-1}$ (red points of Fig.~\ref{fig:bmass_fixedvel}). As this `reversed' trend is not seen for the spectral line widths (see below), it is possible that there is a redshift evolution between $V_{\rm flat}$ and $\sigma_{\rm DM}$ which does not occur for the spectral line widths with $\sigma_{\rm DM}$. We investigate this in Section~\ref{sec:vel_dmvd}. We also note that, as in the simulated galaxy samples in \cite{Brook2016} and \cite{Sales2017}, there is a turn-off in the BTFR at the lower mass or rotational velocity end ($V$~$<$~100~km\,s$^{-1}$) for our simulated \simba\ galaxies for $V_{\rm flat}$, which hence affects the BTFR linear fits on which the trends in Fig.~\ref{fig:bmass_fixedvel} rely upon. 

Considering the spectral line width measures now, $W_{\rm 20}$/2 has the same general trend as $\sigma_{\rm DM}$ in Fig.~\ref{fig:bmass_fixedvel}, with the higher redshift disk galaxies having a lower baryonic mass at fixed velocity ($3\sigma$ significance) than the lower redshift disk galaxies. This is once again opposite to the distribution seen visually for $V_{\rm flat}$, although all observational measures show a steepening in the slope between redshift 0 and 1 (second half of Table~\ref{tab:btfr_fits}). While the scatter remains largest for $W_{\rm 50}$/2, and there is no clear redshift trend in the bottom panel of Fig.~\ref{fig:bmass_fixedvel} (green points), at 100~km\,s$^{-1}$ the corresponding baryonic mass from the best fit relations follows the same trend as for $\sigma_{\rm DM}$ and $W_{\rm 20}$. $W_{\rm 50}$ also visually agrees with that of $W_{\rm 20}$/2 and $\sigma_{\rm DM}$ in that higher redshift sources lie on the lower right of the BTFR plot. Quantitatively, both the slope and y-intercept at $z$ = 1 for $W_{\rm 20}$ differ from the lower redshift samples by over 3$\sigma$, and this is also seen for $W_{\rm 50}$/2. We note that source numbers are lower for this disky subsample (Table~\ref{tab:btfr_fits}), but that these numbers are again comparable to the expected number from the LADUMA survey. 

The above analysis depends on the best BTFR fits made to each of these relations, and for many there is no clear trend in slope or intercept with redshift for each sample from one redshift sample to another, within the total redshift range of 0 to 1. As described in the following subsection, studies such as \cite{Ubler2017} suggest that the zero-point, or y-intercept, of the BTFR changes between redshifts $z$~=~0 to 0.9. To see if this trend exists for our simulated LADUMA sample, in Fig.~\ref{fig:btfr_redshift_fixedslope} we repeat our BTFR fits but fix the slope in each case to the slope found in the redshift $z$~=~0 case (Table~\ref{tab:btfr_fits}). For $\sigma_{\rm DM}$ and the spectral line widths, we see a decrease in the y-intercept of these best-fit relations with increasing redshift, which again is seen to a $\sim3\sigma$ significance. The trend is once again reversed for $V_{\rm flat}$. While we stress that the effect is weak, especially for instance between redshifts $z$~=~0 and 0.75 in the case of $W_{\rm 20}$, there is evidence of redshift evolution in the y-intercept for this simulated LADUMA sample when the slope is fixed.


We hence conclude that LADUMA will be able to study and detect a weak redshift evolution of the BTFR through the use of spectral line widths. Such a study will be better suited to using $W_{\rm 20}$/2 rather than $W_{\rm 50}$/2, and will benefit from focusing on disky galaxies identified through auxiliary data to best observe this effect. As LADUMA will already be relying on additional, non-radio data available to e.g. obtain stellar masses necessary for this study, this is encouraging news.


There are a couple of caveats to this. For one, while our \simba\ H{\sc i} data cubes are already convolved with a radio beam and with noise included in the spectrum, all galaxies were rotated so that no inclination correction was required. This correction effect and potential difficulty in obtaining accurate inclination measures for high redshift targets, assumptions in the stellar mass-to-light ratio used for the baryonic mass measure, and any additional observing factors such as RFI, will introduce errors into the study to be undertaken in the LADUMA survey. Lastly, there remains discrepancies between \simba\ simulated galaxies and true galaxies, such as an excess amount of dark matter in the core of \simba\ galaxies \citep{Glowacki2020}. Nonetheless, we expect that such factors will not prevent the successful study of the H{\sc i} redshift evolution of the BTFR in LADUMA.

\subsubsection{Comparison with previous studies}\label{sec:btfr_comp}

We briefly compare our findings with other simulated and observational studies mentioned in the introduction. We note, as summarised in \cite{Abril-Melgarejo2021}, that the evolution of the TFR with cosmic time has remained a matter of debate, and that many different methods (e.g. integral field spectroscopy) to study the TFR have been used, as opposed to H{\sc i} observations from radio telescopes that we simulate here. 

\cite{Portinari2007} used an improved version of `SGP03' simulations \citep{Sommer-Larsen2003}, and found negligible redshift evolution between a redshift of $z$~=~0--1 in the \emph{stellar} TFR, although their sample only included $\sim$25 disk galaxies. Their findings with simulated galaxies for the stellar TFR agrees with various observational studies, such as \cite{Tiley2019} between $z~=~$0 and 1. However, redshift evolution of the zero point (i.e. the y-intercept) of the stellar TFR was found in \cite{Puech2008} between redshifts of $z~=~$0 and 0.6, and \cite{Ubler2017} found the same for redshifts between 0 and 0.9, although not between redshifts of 0.9 to 2.3. \cite{Barden2003}, through near-infrared H$\alpha$ observations of $z~\approx~$1 disc galaxies, found a brightening by 1.1~mag in the rest-frame B-band TFR compared to local galaxies, while \cite{Bamford2006} saw a brightening of 1.0~mag between $z~=~$0 and 1 spiral galaxies. 

We highlight that the inclusion of the \HI\ mass, as well as the approach in measuring the TFR using \HI\ linewidths, is an important difference between our work on the BTFR and studies focusing on the stellar TFR. One example of this is that low-mass systems are completely H{\sc i}-dominated in terms of baryonic content. Nonetheless, the trend \cite{Puech2008} and \cite{Ubler2017} observe in the stellar TFR agrees with what we find in Fig.~\ref{fig:btfr_redshift_krot} for the BTFR: at comparable stellar or baryonic masses, higher redshift galaxies rotate faster. This can be explained through the virial theorem -- since $M$~$\propto$ $RV^{2}$, then for an higher redshift galaxy with the same mass $M$ and smaller radius $R$ than a local galaxy, the velocity $V$  of the higher redshift galaxy must hence increase.  In \simba, as in observations, the size of star-forming galaxies decreases to higher redshifts~\citep{Appleby2020}, thereby driving a higher rotational velocity.

\cite{Obreschkow2009} consider the BTFR through the Millennium Simulation for 3$\times$10$^{7}$ galaxies of various morphologies, albeit at a far broader range of redshifts ($z$~=~0, 4.89, and 10.07). They use the $W_{\rm 20}$ measure of the H{\sc i} content and find that galaxies of identical baryonic mass have broader H{\sc i} profiles, and hence larger circular velocities, at higher redshift, a conclusion we find agreement with for \simba\ galaxies. \cite{Ubler2017} also find a change in the zero-point BTFR between observed galaxies at $z~=~$0 and $z~=~0.9$, which compares well to our finding with $\sigma_{\rm DM}$, and to a lesser extent with $W_{\rm 20}$/2, where we had fixed the slope (Fig.~\ref{fig:btfr_redshift_fixedslope}). 

Conversely, \cite{Puech2010} did not see an evolution in either slope or zero-point/y-intercept in the BTFR through VLT (Very Large Telescope) observations between redshifts of 0.0--0.6. This roughly matches our finding for spiral galaxies between redshifts of 0.0--0.5 through $W_{\rm 20}$/2, where the slope and y-intercept best-fit parameters agree within 1$\sigma$. It may hence mean that the LADUMA study of the redshift evolution of the BTFR will only become significantly apparent at higher redshifts, i.e. $z~\sim~$1.

\cite{Obreschkow2009} also noted that the scatter of their relation increases with redshift, which they attribute to a lower degree of virialisation at higher redshifts in their model. The increased scatter was also found in the observational study of \cite{Puech2010}, who attributed this to major merger events for their higher redshift sample. We see the same effect in increasing scatter for both $W_{\rm 50}$/2 and $W_{\rm 20}$/2 in the full sample with increasing redshift, noting however here we have lower sample sizes and a much smaller redshift range than the sample of \cite{Obreschkow2009}, and hence observe this trend at low significance. While we see an increase in scatter for $\sigma_{\rm DM}$ between redshifts of 0 to 0.5, the scatter decreases for the redshift $z~=~1$ galaxies - possibly an effect of lower sample size, or perhaps a lower fraction of major mergers for these galaxies than in lower redshift snapshots here, despite our efforts to remove merging or interacting galaxies from our samples through visual inspection.

\subsection{Rotational velocity with $\sigma_{\rm DM}$}\label{sec:vel_dmvd}

As introduced before, we also wish to consider how well do the three observational measures for rotational velocity compare with a key attribute of galaxies that \emph{cannot} be measured from observational studies, the velocity dispersion of the dark matter particles associated with the host galaxy ($\sigma_{\rm DM}$). As this information is available within \simba\, we can investigate this directly, both for the full sample, when considering the disky subsample, and at different redshifts. Given that the observational velocity measures have been used with great success in studying the various forms of the TFR in the literature, and shows general agreement in this study, we expect observational measures to correlate with $\sigma_{\rm DM}$. However, as the dark matter halo properties were found to change with redshift in \simba\ galaxies, as demonstrated in \citet{Dave2019}, it is important to investigate if these correlations evolve with redshift, as this could affect our interpretation of the BTFR redshift evolution results, such as the reverse trend seen for $V_{\rm flat}$ compared to other measures. This is also important to consider for mass modelling of galaxies, which will be done for LADUMA and could be studied as a function of redshift.

\begin{figure*}
\centering
\begin{subfigure}[b]{0.99\textwidth}
  \includegraphics[width=1.\linewidth]{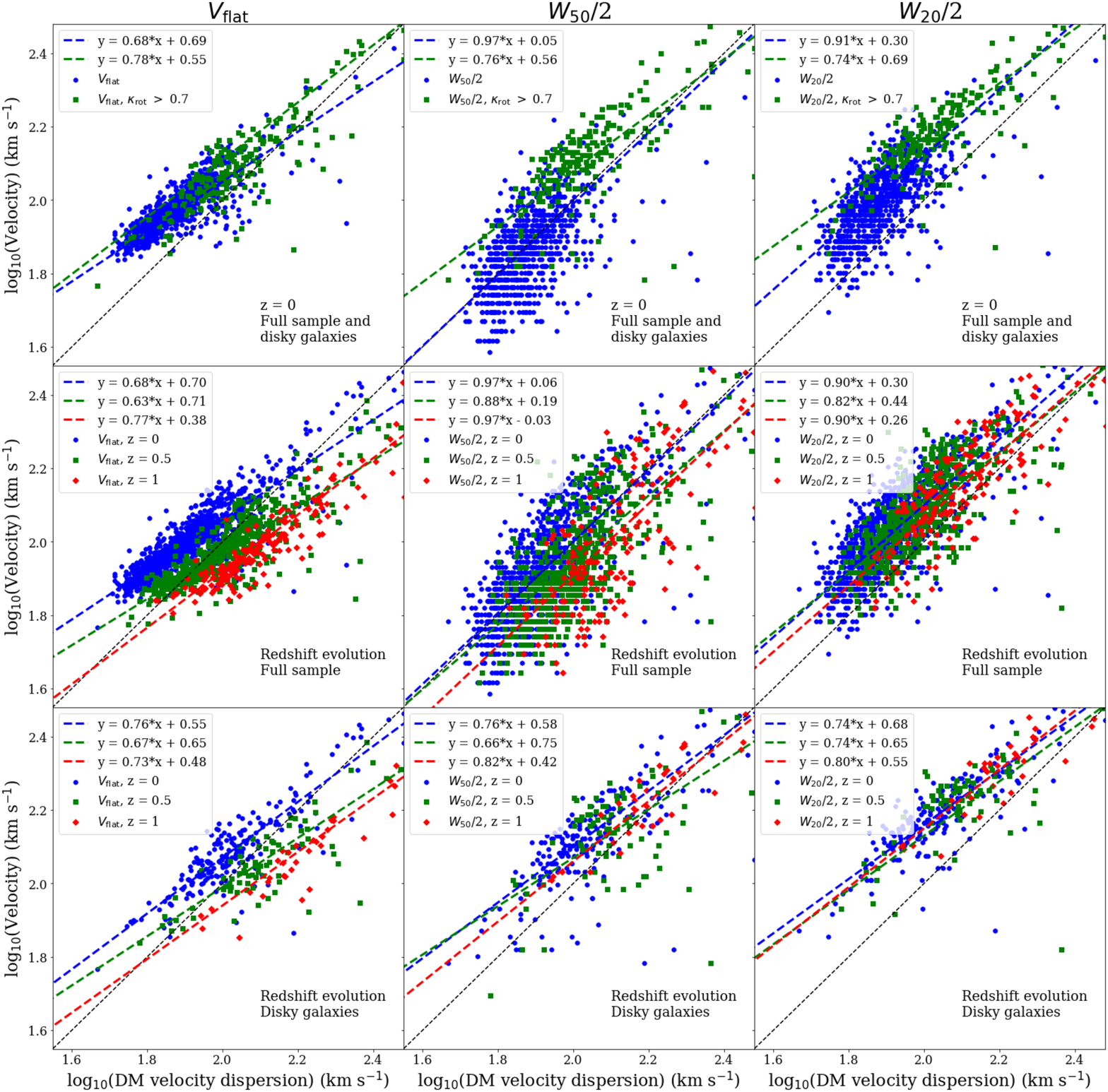}
\end{subfigure}
\caption{A comparison of log$_{\rm 10}$($V$) vs. log$_{\rm 10}$($\sigma_{\rm DM}$). For log$_{\rm 10}$($V$), we test the rotational velocity of \simba\ galaxies calculated from rotation curves through $V_{\rm flat}$ (left panels), and from the H{\sc i} spectra line profiles ($W_{\rm 50}$/2 and $W_{\rm 20}$/2, middle and right panels respectively. These are plotted against the dark matter velocity dispersion ($\sigma_{\rm DM}$) from the \simba\ snapshot properties as calculated in \caesar. The 1:1 trend is given in the black dashed line for scale.\\
\emph{Top row}: At redshift $z$~=~0 we consider the full sample from \simba-hires in blue circles, and the `disky' galaxy subsample ($\kappa_{\rm rot}$~$>$~0.7) in green squares, which tends to pick faster rotating galaxies. Of the three measures here, $V_{\rm flat}$ gives the tightest correlation, while $W_{\rm 50}$/2 is closest to the 1:1 line but has larger scatter, particularly for lower mass galaxies. \\
\emph{Middle row}: We show the redshift evolution for the full sample. Blue circles indicate $z$~=~0 galaxies, green squares for $z$~=~0.5, and red diamonds for $z$~=~1.\\
\emph{Bottom row}: The redshift evolution for disky galaxies only. In both samples, redshift evolution is only seen from the $V_{\rm flat}$ measure.}
\label{fig:vrot_vs_dmvd}
\end{figure*}

\begin{table*}
    \centering
    \begin{tabular}{lccccc}
    \hline 
Velocity definition & Redshift & Sample size & $m$ & $b$ & $\sigma_{\perp}$\\
\hline
Full sample\\
\hline
$V_{\rm flat}$ & 0 & 1043 & 0.68$\pm$0.01 & 0.70$\pm$0.02 & 0.039$\pm$0.001\\
$V_{\rm flat}$ & 0.5 & 559 & 0.63$\pm$0.02 & 0.71$\pm$0.03 & 0.044$\pm$0.001\\
$V_{\rm flat}$ & 1 & 180 & 0.77$\pm$0.04 & 0.38$\pm$0.08 & 0.036$\pm$0.005\\
\hline
$W_{\rm 50}$/2 & 0 & 1043 & 0.97$\pm$0.03 & 0.06$\pm$0.05 & 0.073$\pm$0.003\\
$W_{\rm 50}$/2 & 0.5 & 559 & 0.88$\pm$0.03 & 0.19$\pm$0.06 & 0.084$\pm$0.006\\
$W_{\rm 50}$/2 & 1 & 180 & 0.97$\pm$0.05 & -0.03$\pm$0.10 & 0.088$\pm$0.011\\
\hline
$W_{\rm 20}$/2 & 0 & 1043 & 0.90$\pm$0.02 & 0.30$\pm$0.03 & 0.059$\pm$0.002\\
$W_{\rm 20}$/2 & 0.5 & 559 & 0.82$\pm$0.03 & 0.44$\pm$0.06 & 0.059$\pm$0.002\\
$W_{\rm 20}$/2 & 1 & 180 & 0.90$\pm$0.03 & 0.26$\pm$0.07 & 0.061$\pm$0.003\\
\hline
Disky galaxy subsample ($\kappa_{\rm rot}$~$>$~0.7)\\
\hline
$V_{\rm flat}$ & 0 & 204 & 0.76$\pm$0.02 & 0.55$\pm$0.03 & 0.049$\pm$0.002\\
$V_{\rm flat}$ & 0.5 & 86 & 0.67$\pm$0.03 & 0.65$\pm$0.06 & 0.057$\pm$0.005\\
$V_{\rm flat}$ & 1 & 34 & 0.73$\pm$0.05 & 0.48$\pm$0.10 & 0.041$\pm$0.007\\
\hline
$W_{\rm 50}$/2 & 0 & 204 & 0.76$\pm$0.03 & 0.58$\pm$0.06 & 0.073$\pm$0.002\\
$W_{\rm 50}$/2 & 0.5 & 86 & 0.66$\pm$0.04 & 0.75$\pm$0.09 & 0.103$\pm$0.009\\
$W_{\rm 50}$/2 & 1 & 34 & 0.82$\pm$0.05 & 0.42$\pm$0.10 & 0.047$\pm$0.005\\
\hline
$W_{\rm 20}$/2 & 0 & 204 & 0.74$\pm$0.03 & 0.68$\pm$0.05 & 0.052$\pm$0.002\\
$W_{\rm 20}$/2 & 0.5 & 86 & 0.74$\pm$0.04 & 0.65$\pm$0.08 & 0.063$\pm$0.005\\
$W_{\rm 20}$/2 & 1 & 34 & 0.80$\pm$0.03 & 0.55$\pm$0.06 & 0.044$\pm$0.003\\
\hline
    \end{tabular}
    \caption{Best log-log fits between the three rotational velocity measures and the dark matter velocity dispersion ($\sigma_{\rm DM}$), as in Table~\ref{tab:btfr_fits} (see Fig.~\ref{fig:vrot_vs_dmvd}). Values for the different redshift snapshots and the full sample or disky subsample are given. The disky subsample has similar values to the full sample when using $V_{\rm flat}$. The largest scatter is typically seen for the $W_{\rm 50}$/2 measure.}
    \label{tab:vel_dmvd_fits}
\end{table*}

\begin{table*}
    \centering
    \begin{tabular}{lccccc}
    \hline 
Velocity definition & Redshift & Sample size & $m$ & $b$ & $\sigma_{\perp}$\\
\hline
Full sample\\
\hline
$V_{\rm flat}$ & 0 & 1043 & 0.75$\pm$0.01 & 0.80$\pm$0.01 & 0.027$\pm$0.001\\
$V_{\rm flat}$ & 0.5 & 559 & 0.78$\pm$0.02 & 0.57$\pm$0.03 & 0.031$\pm$0.002\\
$V_{\rm flat}$ & 1 & 180 & 0.71$\pm$0.03 & 0.59$\pm$0.06 & 0.039$\pm$0.012\\
\hline
$W_{\rm 50}$/2 & 0 & 1043 & 1.04$\pm$0.02 & 0.24$\pm$0.03 & 0.064$\pm$0.003\\
$W_{\rm 50}$/2 & 0.5 & 559 & 0.99$\pm$0.03 & 0.15$\pm$0.05 & 0.077$\pm$0.004\\
$W_{\rm 50}$/2 & 1 & 180 & 0.97$\pm$0.06 & 0.08$\pm$0.11 & 0.095$\pm$0.013\\
\hline
$W_{\rm 20}$/2 & 0 & 1043 & 0.98$\pm$0.01 & 0.46$\pm$0.02 & 0.046$\pm$0.001\\
$W_{\rm 20}$/2 & 0.5 & 559 & 0.97$\pm$0.02 & 0.33$\pm$0.04 & 0.054$\pm$0.003\\
$W_{\rm 20}$/2 & 1 & 180 & 0.99$\pm$0.03 & 0.20$\pm$0.07 & 0.061$\pm$0.005\\
\hline
Disky galaxy subsample ($\kappa_{\rm rot}$~$>$~0.7)\\
\hline
$V_{\rm flat}$ & 0 & 204 & 0.79$\pm$0.01 & 0.74$\pm$0.02 & 0.030$\pm$0.002\\
$V_{\rm flat}$ & 0.5 & 86 & 0.87$\pm$0.02 & 0.42$\pm$0.05 & 0.027$\pm$0.001\\
$V_{\rm flat}$ & 1 & 34 & 0.74$\pm$0.05 & 0.56$\pm$0.10 & 0.032$\pm$0.008\\
\hline
$W_{\rm 50}$/2 & 0 & 204 & 0.77$\pm$0.02 & 0.80$\pm$0.03 & 0.062$\pm$0.002\\
$W_{\rm 50}$/2 & 0.5 & 86 & 0.83$\pm$0.06 & 0.59$\pm$0.11 & 0.081$\pm$0.008\\
$W_{\rm 50}$/2 & 1 & 34 & 0.73$\pm$0.05 & 0.75$\pm$0.11 & 0.049$\pm$0.016\\
\hline
$W_{\rm 20}$/2 & 0 & 204 & 0.78$\pm$0.01 & 0.85$\pm$0.02 & 0.035$\pm$0.001\\
$W_{\rm 20}$/2 & 0.5 & 86 & 0.81$\pm$0.03 & 0.69$\pm$0.07 & 0.063$\pm$0.006\\
$W_{\rm 20}$/2 & 1 & 34 & 0.74$\pm$0.03 & 0.79$\pm$0.06 & 0.040$\pm$0.005\\
\hline
    \end{tabular}
    \caption{As in Table~\ref{tab:vel_dmvd_fits}, but for the dark matter velocity dispersion when considering dark matter particles out to a distance determined by the H{\sc i} mass-size relation (Fig.~\ref{fig:vel_vs_DMVD2}).}
    \label{tab:vel_dmvdhi_fits}
\end{table*}

In the nine-panel plot given in Fig.~\ref{fig:vrot_vs_dmvd}, we compare the log$_{\rm 10}$ rotational velocity measures with the logarithm of $\sigma_{\rm DM}$ from the associated dark matter halo (i.e. log$_{\rm 10}$($V$) vs. log$_{\rm 10}$($\sigma_{\rm DM}$)). The left column gives $V_{\rm flat}$, the middle column $W_{\rm 50}$/2, and the right column $W_{\rm 20}$/2. The first row focuses on $z$~=~0 galaxies, comparing the full sample (blue) with the disky subsample (green). The middle row considers the full sample at each redshift bin: blue circles indicate $z$~=~0 galaxies, green squares for $z$~=~0.5, and red diamonds for $z$~=~1. The bottom row repeats this, but gives the redshift evolution of the $V$-$\sigma_{\rm DM}$ relation for disky galaxies only. The 1:1 line is given in the black dashed line for reference. The best linear fit parameters and scatter measure is given in Table~\ref{tab:vel_dmvd_fits} for each measure and redshift, separated into the full sample and disky galaxy subsample. We note that we consider both the change in slope and normalisation (y-intercept) of the relations in regards to redshift evolution in the following analysis.

First, we compare how each observational rotation measure compares in tracing $\sigma_{\rm DM}$. At all redshifts, $V_{\rm flat}$ gives the tightest correlation (least scatter) with $\sigma_{\rm DM}$. The smaller scatter may be attributed to the fact the rotation curves, which $V_{\rm }$ is measured from, considers the distribution of all particles in a \simba\ galaxy and its halo, while spectral line widths only come from the H{\sc i} gas particle information. As the H{\sc i} content can be pressure supported, these measures can give different results for slow-rotating galaxies at the bottom end of the BTFR. $W_{\rm 20}$/2 unsurprisingly once again performs better than $W_{\rm 50}$/2 in regards to scatter - but we note that for the full sample $W_{\rm 50}$/2 best matches the 1:1 relation. Therefore, $W_{\rm 20}$/2 may be better suited for only disky galaxies when comparing it with $\sigma_{\rm DM}$.

In the full sample at redshift $z$~=~0, $V_{\rm flat}$ gives the flattest slope and highest y-intercept, by over 5$\sigma$, compared to the spectral line measures. The spectral line width measures give a steeper, close-to-unity slope. However, in the disky subsample the $V$-$\sigma_{\rm DM}$ best-fit slopes (values around 0.75) and y-intercepts found from all observation methods agree within 1$\sigma$ at identical redshifts. Therefore, each observational method correlates with the dark matter velocity dispersion in the same manner for rotation-dominated galaxies. As $V_{\rm flat}$ is only usable to low-redshift due to spatial resolution requirements, the spectral line widths measures would be more useful for LADUMA in this instance.

Next, we focus on the differences seen between the disky and full populations. Relative to the full sample, disky galaxies have higher $\sigma_{\rm DM}$ measures, and also higher rotational velocities. Quantitatively, the differences between the samples are more evident when using spectral line widths. The best-fitting slopes for the disky subsample and the full sample at all redshifts for $V_{\rm flat}$ agree within 2$\sigma$, as do their intercepts (sometimes within 1$\sigma$). This is not the case for the spectral line widths (top row of Fig.~\ref{fig:vrot_vs_dmvd}) - disky galaxies have flatter slopes and correspondingly higher y-intercepts than the full sample. Therefore, spectral line widths better trace the effect of galaxy morphology in its correlation with $\sigma_{\rm DM}$ from the full dark matter halo, than velocity measures from galaxy rotation curves. 

We now consider the redshift evolution of the $V_{\rm obs}$-$\sigma_{\rm DM}$ relation. At higher redshifts, galaxies have a lower rotational velocity at fixed $\sigma_{\rm DM}$. This redshift evolution is diminished for the disky subsample, and despite persisting for $V_{\rm flat}$ is essentially negligible when considering spectral line width measures. Redshift evolution of this relation for the full sample, which includes dispersion-dominated galaxies, might be attributed to a different merger history compared to the disky galaxies - that is, at higher redshifts dispersion-dominated galaxies rotate slower than they do now, while this is not so apparent when considering spiral galaxies. Alternatively, at fixed rotational velocities, there is a lower dark matter velocity dispersion, which is tied to a smaller dark matter halo mass - so hence at higher redshift galaxies have a smaller dark matter halo, which is more apparent for dispersion-dominated galaxies. However in Section~\ref{sec:vel_DMHM}, we note little to no redshift evolution between $V_{\rm obs}$ and dark matter halo mass.

\begin{figure*}
\centering
\begin{subfigure}[b]{0.99\textwidth}
  \includegraphics[width=1.\linewidth]{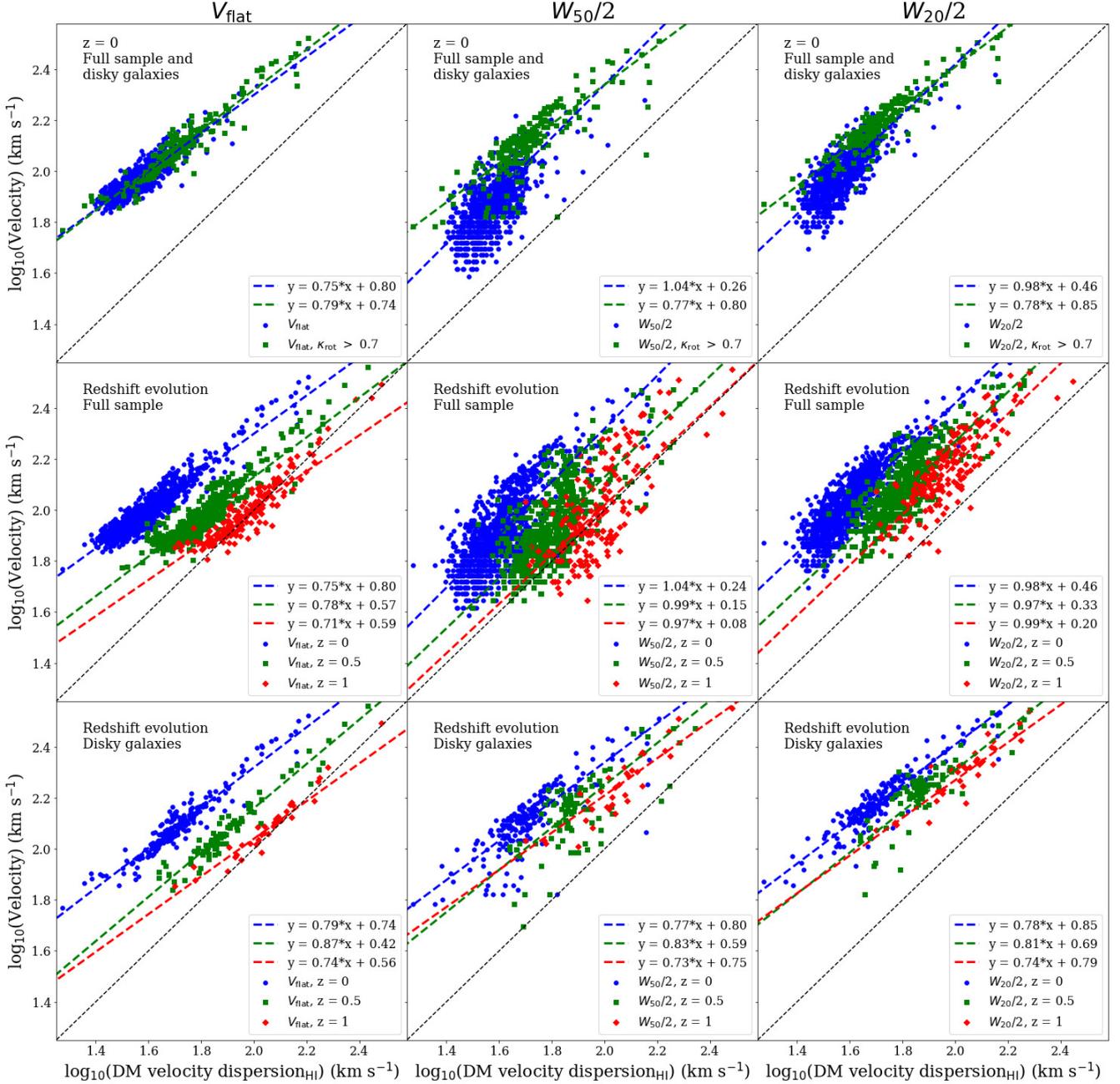}
\end{subfigure}
\caption{As in Fig.~\ref{fig:vrot_vs_dmvd}, but using the dark matter velocity dispersion calculated from particles within a radius from the galaxy centre determined by the H{\sc i} mass-size relation - therefore, lower dispersion values are seen. Redshift evolution also stronger here, and the correlation slightly tighter at z~=~0 and 0.5, but the relation shifts from the 1:1 line.}
\label{fig:vel_vs_DMVD2}
\end{figure*}

\begin{figure}
\centering
\begin{subfigure}[b]{0.49\textwidth}
  \includegraphics[width=0.95\linewidth]{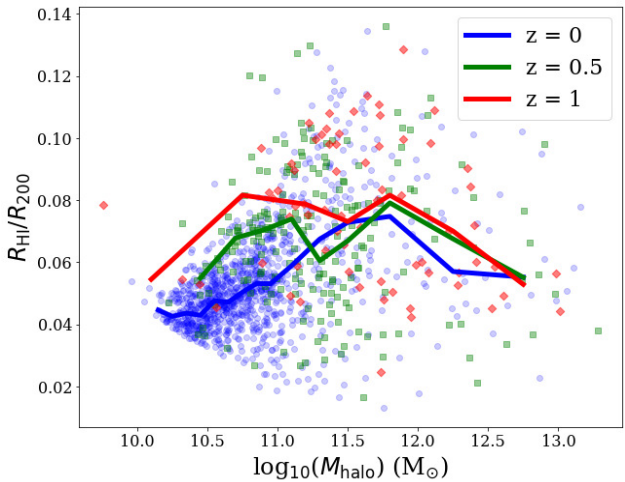}
\end{subfigure}%
\caption{The ratio of the H{\sc i} radius ($R_{\rm HI}$) determined by the H{\sc i} galaxy masses and the relation in \citet{Wang2016}, and $R_{\rm 200}$, versus total halo mass for \simba\ galaxies in each redshift snapshot. The running median for each redshift is given by the bold lines.}
\label{fig:radiuscomp}
\end{figure}

\begin{table*}
    \centering
    \begin{tabular}{lccccc}
    \hline 
Velocity definition & Redshift & Sample size & $m$ & $b$ & $\sigma_{\perp}$\\
\hline
Full sample\\
\hline
$V_{\rm flat}$ & 0 & 1043 & 0.15$\pm$0.01 & 0.35$\pm$0.03 & 0.057$\pm$0.001\\
$V_{\rm flat}$ & 0.5 & 559 & 0.17$\pm$0.01 & 0.14$\pm$0.05 & 0.053$\pm$0.001\\
$V_{\rm flat}$ & 1 & 179 & 0.16$\pm$0.02 & 0.16$\pm$0.09 & 0.069$\pm$0.003\\
\hline
$W_{\rm 50}$/2 & 0 & 1043 & 0.26$\pm$0.01 & -0.85$\pm$0.05 & 0.095$\pm$0.001\\
$W_{\rm 50}$/2 & 0.5 & 559 & 0.25$\pm$0.01 & -0.79$\pm$0.07 & 0.102$\pm$0.001\\
$W_{\rm 50}$/2 & 1 & 179 & 0.25$\pm$0.02 & -0.86$\pm$0.16 & 0.132$\pm$0.002\\
\hline
$W_{\rm 20}$/2 & 0 & 1043 & 0.23$\pm$0.01 & -0.50$\pm$0.04 & 0.071$\pm$0.001\\
$W_{\rm 20}$/2 & 0.5 & 559 & 0.24$\pm$0.01 & -0.55$\pm$0.06 & 0.069$\pm$0.001\\
$W_{\rm 20}$/2 & 1 & 179 & 0.22$\pm$0.03 & -0.30$\pm$0.12 & 0.093$\pm$0.001\\
\hline
$W_{\rm 20}$/2 & All & 1782 & 0.24$\pm$0.01 & -0.52$\pm$0.05 & 0.074$\pm$0.001\\
\hline
Disky galaxy subsample ($\kappa_{\rm rot}$~$>$~0.7)\\
\hline
$V_{\rm flat}$ & 0 & 204 & 0.19$\pm$0.01 & -0.05$\pm$0.09 & 0.064$\pm$0.002\\
$V_{\rm flat}$ & 0.5 & 86 & 0.18$\pm$0.01 & -0.03$\pm$0.09 & 0.065$\pm$0.008\\
$V_{\rm flat}$ & 1 & 33 & 0.20$\pm$0.02 & -0.21$\pm$0.21 & 0.059$\pm$0.007\\
\hline
$W_{\rm 50}$/2 & 0 & 204 & 0.19$\pm$0.01 & -0.08$\pm$0.13 & 0.087$\pm$0.002\\
$W_{\rm 50}$/2 & 0.5 & 86 & 0.20$\pm$0.02 & -0.09$\pm$0.19 & 0.110$\pm$0.007\\
$W_{\rm 50}$/2 & 1 & 33 & 0.23$\pm$0.02 & -0.43$\pm$0.20 & 0.067$\pm$0.014\\
\hline
$W_{\rm 20}$/2 & 0 & 204 & 0.19$\pm$0.01 & 0.07$\pm$0.08 & 0.059$\pm$0.001\\
$W_{\rm 20}$/2 & 0.5 & 86 & 0.20$\pm$0.01 & -0.04$\pm$0.16 & 0.088$\pm$0.007\\
$W_{\rm 20}$/2 & 1 & 33 & 0.23$\pm$0.03 & -0.38$\pm$0.29 & 0.068$\pm$0.016\\
\hline
$W_{\rm 20}$/2 & All & 323 & 0.21$\pm$0.01 & -0.12$\pm$0.12 & 0.073$\pm$0.003\\
\hline
    \end{tabular}
    \caption{As in Table~\ref{tab:vel_dmvd_fits}, but for the dark matter halo mass (Fig.~\ref{fig:vel_vs_DMHM}). We add additional best-fit parameters for $W_{\rm 20}$/2 across all three redshift snapshots, for the full sample and disky subsample.}
    \label{tab:vel_dmhm_fits}
\end{table*}

\begin{figure*}
\centering
\begin{subfigure}[b]{0.99\textwidth}
      \includegraphics[width=1.\linewidth]{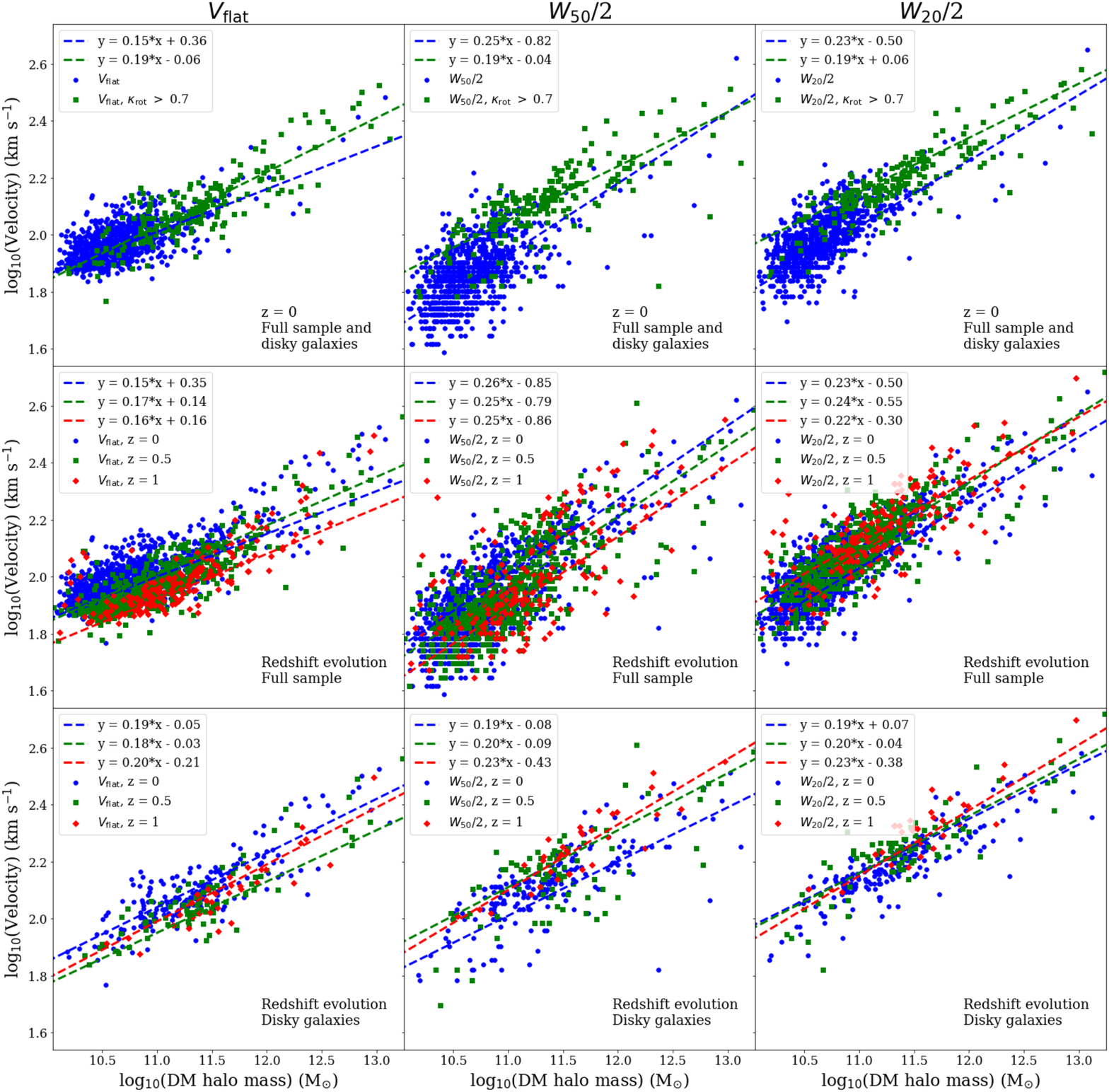}
\end{subfigure}
\caption{As in Fig.~\ref{fig:vrot_vs_dmvd}, but plotting rotational velocity against the dark matter halo mass for the entire galaxy. The deviation from the best fitting line is evident for $W_{\rm 50}$/2 at lower masses. Only weak redshift evolution can be seen.}
\label{fig:vel_vs_DMHM}
\end{figure*}

When considering the full sample quantitatively, the slope does not change between redshift 0 and 1 beyond a 2--3$\sigma$ level. Meanwhile, the y-intercept $b$ changes more significantly (over 3$\sigma$) for $V_{\rm flat}$ with redshift. The y-intercept does not significantly evolve with redshift when probed by spectral line width measures, beyond being slightly higher at $z~=~0.5$ by 2$\sigma$. 

For disky galaxies, there is only an evolution of $V_{\rm flat}$ at fixed $\sigma_{\rm DM}$ with redshift (higher rotational velocity at lower redshift) - not for the spectral line widths. This can explain the `reverse trend' we see in Fig.~\ref{fig:bmass_fixedvel}, where at fixed velocity the baryonic mass for $\sigma_{\rm DM}$ and $W_{\rm 20}$ decreases with increasing redshift, but increases for $V_{\rm flat}$. The lack of redshift evolution of $W_{\rm 20}$ with $\sigma_{\rm DM}$, meanwhile, suggests that there is no additional complication in using $W_{\rm 20}$ for LADUMA in studying the redshift evolution of the BTFR.


We remind the reader that the dark matter velocity dispersion measurements come from the dark matter halo that the host galaxy is associated with. This hence means that the gas content traced by spectral line widths will not completely correlate with the attributes of the dark matter halo. What if we consider the dark matter particles within the baryonic extent of the galaxy, rather than the full galaxy halo?

\subsection{Velocity dispersion of dark matter within the baryonic extent}

It will be important to investigate if there is a distinct difference between observational velocity and the dispersion measure for dark matter within the same extent as the baryonic matter, and the dispersion measure for the full dark matter halo. For example, with mass modelling of galaxies, the dark matter content and distribution is vital to model properly at all distances from the centre, including at inner radii. Furthermore, we can investigate if there will be any redshift evolution of the dark matter content in the inner parts of galaxies, which can inform us on galaxy morphological evolution with redshift.

We repeat the same panel scheme as before in Fig.~\ref{fig:vel_vs_DMVD2}, but this time give the dark matter velocity dispersion for particles occupying the same radius from the galaxy centre of mass determined by the H{\sc i} mass-size relation, using that found observationally in \cite{Wang2016}. This is the same radius we extend the galaxy rotation curves generated to obtain the $V_{\rm flat}$ measure. The key difference between this and Fig.~\ref{fig:vrot_vs_dmvd} is that now we only consider the dark matter content within the same extent that would be traced observationally in LADUMA (i.e. the same region as the H{\sc i} gas), not the content of the entire dark matter halo of the galaxy. 

In Fig.~\ref{fig:radiuscomp} we show that for most galaxies the radius we extend to in Fig.~\ref{fig:vel_vs_DMVD2} ($R_{\rm HI}$) is less than 1/10$^{\rm th}$ of $R_{\rm 200}$, which corresponds to the radius enclosing the total halo mass for \simba\ galaxies, irrespective of total galaxy halo mass $M_{\rm halo}$. Table~\ref{tab:vel_dmvdhi_fits} gives the best fit parameters for the relation explored in Fig.~\ref{fig:vel_vs_DMVD2}, henceforth $V$-$\sigma_{\rm DM-HI}$.

Many of the same findings from Fig.~\ref{fig:vrot_vs_dmvd} and Table~\ref{tab:vel_dmvd_fits} still hold; namely that $V_{\rm flat}$-$\sigma_{\rm DM-HI}$ gives the tightest relation and $W_{\rm 50}$/2 the least, and that the spectral line widths trace different values of $\sigma_{\rm DM}$ for disky galaxies compared to other morphologies. However, there are a few differences - for instance, for the full sample at all redshifts, the spectral line widths give a slope much better matching a value of unity in $V$-$\sigma_{\rm DM-HI}$. This shows a stronger correlation between these measures, to be expected when they consider the same volume. Likewise, the scatter is generally slightly smaller when considering $\sigma_{\rm DM-HI}$. 

One obvious new effect when comparing $V$-$\sigma_{\rm DM}$ with $V$-$\sigma_{\rm DM-HI}$ is that in the latter measure, all points shift away (toward the upper left of the plot-space) from the 1:1 relation line; that is, they have lower dark matter velocity dispersion measures when only considering $\sigma_{\rm DM-HI}$. This particular result is entirely expected, as the velocity dispersion should naturally decrease when only considering a subset of dark matter particles in the galaxy halo. 

The second effect, which was not expected, is that redshift evolution is much more pronounced, even for disky galaxies when considering spectral line width measures. This is both evident visually and when considering the y intercept for the full sample between $z~=~0$ and 1, which can differ by $\sim3\sigma$. This suggests that there is some difference in the dark matter distribution within the baryon extent in \simba\ galaxies based on host galaxy morphology, which is less apparent from H{\sc i} spectral line profiles when considering the entire galaxy dark matter halo. Nonetheless, the redshift evolution in the y-intercept remains strongest when using $V_{\rm flat}$. The slope does not evolve.

\subsection{Velocity versus dark matter halo mass}\label{sec:vel_DMHM}

Lastly, we consider how well each velocity measure corresponds with the dark matter halo mass, $M_{\rm DMH}$. As before, we repeat the nine-panel plot for this relation in Fig.~\ref{fig:vel_vs_DMHM}. Once again, $V_{\rm flat}$ has the least scatter but the flattest slope, and therefore is less correlated with $M_{\rm DMH}$ than spectral line witdh measures. Furthermore, spectral line width measures show a bigger distinction between the disky subsample from all galaxies, and at fixed dark matter halo masses, galaxies rotate faster at lower redshifts. These results are consistent with the findings from the previous two figures, and the strong correlation between dark matter velocity dispersion and dark matter halo mass \citep{Zahid2018}. We do however note that this sample, which matches the expected number of galaxies to be detected by LADUMA, and hence designed to be a prediction of what LADUMA will find, is limited in number -- a bigger sample may find a different result. Wider field surveys such as MIGHTEE-HI and WALLABY will better address this, but will also not extend to as high a redshift ($z \approx 1$) as LADUMA.

The relation of $M_{\rm DMH}$ with line width is remarkably redshift-invariant. Any redshift evolution is hard to distinguish from the spectral line width measurements (within 1$\sigma$ in the difference of slope $m$). Noting the lower numbers for the disky subsample, a redshift evolution effect is slightly more apparent for all velocity measures (within 2--3$\sigma$ in slope), albeit still slight, with convergence at the lower mass end when using $W_{\rm 50}$/2 and $W_{\rm 20}$/2. This means that, up to a redshift of $z$~=~1 at least, the H{\sc i} spectral line width can be used to get a reasonable estimate of the dark matter halo mass.  This is particularly valuable since it is difficult to estimate dark matter halo masses in any other way for moderate-mass galaxies at these redshifts.

Table~\ref{tab:vel_dmhm_fits} gives the best-fit parameters for individual redshift snapshots for the full sample and disky subsamples for all velocity measures, as in previous measures. Given the redshift invariance, we additionally give the best-fit parameters for galaxies combined across all three redshift snapshots for $W_{\rm 20}$/2 (chosen due to its reduced scatter compared to $W_{\rm 50}$/2), for the full sample:

\begin{equation}
    {\rm log_{10}}(W_{\rm 20}/2) = 0.24 {\rm log_{10}}(M_{\rm DMH}) - 0.52,
\end{equation}

where $W_{\rm 20}$/2 is in units of km\,s$^{-1}$ and $M_{\rm DMH}$ in M$_{\odot}$; and the disky subsample:

\begin{equation}
    {\rm log_{10}}(W_{\rm 20}/2) = 0.21 {\rm log_{10}}(M_{\rm DMH}) - 0.12.
\end{equation}

While the slopes are similar between these two samples, the disky subsample has a higher y-intercept (different to about 4$\sigma$). These relations can be used in upcoming H{\sc i} emission surveys across all SKA pathfinder telescopes.

\section{Discussion}\label{sec:discuss}

As established in Section~\ref{sec:intro}, the BTFR provides a tight relation between the mass of the stars and gas with the rotation speed of spiral galaxies. The tightness of the relation across many orders of magnitude suggests at a deep connection between the baryonic content and their dark matter halos. Thus it is important to consider how does this relation evolve with redshift. Redshift evolution of any form of the TFR, as summarised in Section~\ref{sec:btfr_comp} where we compare our results with the literature, has been an area of debate in observational and simulated studies covering various methods of rotational measurement, although studies considering measures such as $W_{\rm 20}$ of H{\sc i} emission profiles in simulated galaxies \citep{Obreschkow2009} did observe redshift evolution of the BTFR.

Previous observational studies specifically using the H{\sc i} 21-cm transition in emission have been limited to lower redshift due to the weakness of the signal. Upcoming H{\sc i} emission surveys with SKA pathfinder telescopes will probe previously unexplored redshift regimes, with LADUMA (the deepest of these surveys) aiming to go redshifts of $z\ga 1$. It is hence expected that LADUMA may be able to present evidence of an evolution of the BTFR, which indeed is one of its science goals. Our study of a sample of simulated galaxies from the \simba\ hydrodynamic simulation, which has similar sample size and mass range of galaxies to the predicted sample from LADUMA, shows that LADUMA will be able to measure such a redshift evolution, although we emphasise the need for LADUMA to discern disky galaxies through ancillary information. 

The $W_{\rm 20}$/2 spectral line width measure shows this evolution most clearly, both in the slope and y-intercept, between $z~=~0$ and 1. $V_{\rm flat}$ also showcases redshift evolution, albeit this measure from rotation curves will not be applicable at high redshift as galaxies will not be spatially resolved, and hence such measures will be impossible. From the spectral line width measures, which agree visually with the trend seen with $\sigma_{\rm DM}$, we find that at equal baryonic masses, higher redshift galaxies rotate faster. This assumes that such galaxies within \simba\ are realistic representations of true galaxies, and that additional effects such as random inclinations of disk galaxies and any potential difficulty in obtaining accurate measures of this and stellar mass-to-light ratios of LADUMA galaxies - see the end of Section~\ref{sec:btfr_disky} for further details. It is also noted that low numbers of disky galaxies are available at higher redshift, resulting in larger errors and a lack of low mass galaxies, hence introducing a bias toward higher mass galaxies. While this sample distribution is expected to occur in LADUMA (Section~\ref{sec:sample}), it is an important point to consider. Larger samples than what will be possible with LADUMA, such as with the SKA itself, or with SKA pathfinder telescopes at lower redshift than LADUMA, may indeed discover different trends if they also probe to lower baryonic masses or greater numbers. The \simba\ simulation also excludes galaxies with stellar masses below our mass resolution limit, although we do not expect many galaxies below this mass limit of 7.25$\times$10$^{8}$~M$_{\odot}$ to be detected in H{\sc i} by LADUMA. 

We also show that with a \simba\ sample chosen to represent LADUMA that there exists a mild redshift evolution of the y-intercept of the relation between the rotational velocity and $\sigma_{\rm DM}$ of the full dark matter halo, while stronger evolution is evident when only considering the $\sigma_{\rm DM}$ of dark matter particles within the same extent as the baryonic matter. Again, other H{\sc i} galaxy samples to LADUMA which probe a different parameter space in redshift and galaxy mass may give a different evolution. A lack of redshift evolution between observational rotational velocity and the dark matter halo mass, meanwhile, showcases the potential of upcoming H{\sc i} emission surveys in estimating the dark matter halo mass, which is a finding we make in addition to our BTFR analysis.

\section{Conclusions}\label{sec:summary}

We have examined three common rotational velocity measures from galaxy rotation curves, using the methodology presented in \cite{Glowacki2020}, and H{\sc i} spectral line widths by creating H{\sc i} data cubes through \martini, for \simba-hires galaxies. With each measure we consider the redshift evolution of the BTFR to predict the results of this study to be undertaken by the LADUMA survey with the MeerKAT array. We then compare each measure with the dark matter velocity dispersion ($\sigma_{\rm DM}$), both from the host galaxy's dark matter halo and particles within the same extent as the baryonic mass ($\sigma_{\rm DM-HI}$), and the dark matter halo mass. The redshift evolution of these trends is hence also predicted for LADUMA.

We find that our spectral line measures follow a 1:1 relation with $V_{\rm flat}$. We note that a downturn in this relation seen for low-mass (or slowly rotating) galaxies is attributed to dispersion-dominated galaxies, which is removed when we only consider the disky subsample. This selection proves crucial in the redshift evolution of the BTFR as well, which observationally only holds for disky galaxies which are kinematically dominated in their rotation. We find a weak redshift evolution of the BTFR between $z$~=~0 to 1 when using $\sigma_{\rm DM}$ and $W_{\rm 20}$ - at fixed velocities the baryonic mass of galaxies decreases with increasing redshift ($z$~=~0 to 1). When fixing the slope from the $z$~=~0 best BTFR fit, the y-intercept decreases with redshift for all measures, except for $V_{\rm flat}$ which instead increases. Spectral line width measures such as $W_{\rm 20}$ are also preferred over narrower measures, e.g. $W_{\rm 50}$, due to reduced scatter. LADUMA will need to identify the morphologies of the host galaxies to best study the redshift evolution of the BTFR.


When comparing observational rotational velocity measures with $\sigma_{\rm DM}$, we find redshift evolution is stronger when using $V_{\rm flat}$ and galaxies of various morphologies, rather than from spectral line widths and the disky galaxy subsample, suggesting the measured $\sigma_{\rm DM}$ change with redshift is dependent on the host galaxy morphology. This redshift evolution from $V_{\rm flat}$ can explain the apparent `reverse' trend found for this observational redshift measure in the BTFR with redshift.

However, the dark matter velocity dispersion of particles within the same extent as the H{\sc i} gas from the galaxy's centre of mass ($\sigma_{\rm DM-HI}$) displays far clearer redshift evolution in both the full and disky sample for all velocity measures. At higher redshifts, galaxies have a lower rotational velocity at fixed $\sigma_{\rm DM}$. $W_{\rm 20}$ again gives a tighter relation here than $W_{\rm 50}$, and while we note $V_{\rm flat}$ is tighter still, this observational measure is not feasible for LADUMA at higher redshifts due to insufficient spatial resolution. 

At most weak redshift evolution is seen between the velocity measures and the dark matter halo mass. From the best fit values, similar slope values $m$ are seen between the full sample and disky subsample, with the y-intercept $b$ being higher in the disky subsample. We give fits for individual redshift snapshots for all rotational velocity measures, and for $W_{\rm 20}$/2 across $z$~=~0--1 from our three snapshots combined for this relation. These fits can hence be used to estimate the dark matter halo mass from H{\sc i} spectral line profile widths with upcoming SKA pathfinder telescope H{\sc i} emission surveys. 

\simba\ has proved instrumental in studying the BTFR, with good comparison with the SPARC survey shown in \cite{Glowacki2020}, and exciting predictions for the LADUMA survey made here. The ability to trace the dynamics of galaxies out to intermediate redshifts using \HI\ spectral linewidths thus provides interesting dynamical information for both galaxies and halos, even without resolving the rotation curves.  As explorations of the \HI\ universe enter into a new era, spectral linewidths promise to provide a key new piece of observational data to explore galaxy evolution and constrain galaxy formation models.


\section*{Acknowledgements}
We thank the anonymous referee for their feedback that have helped improve the paper. We thank Andrew Baker, Sarah Blyth, and Weiguang Cui for useful and helpful discussions. We thank Robert Thompson for developing {\sc Caesar}, and the {\sc yt} team for development and support of {\sc yt}.

RD acknowledges support from the Wolfson Research Merit Award program of the U.K. Royal Society. Ed Elson acknowledges that this research is supported by the South African Radio Astronomy Observatory, which is a facility of the National Research Foundation, an agency of the Department of Science and Technology. Marcin Glowacki acknowledges support from the Inter-University Institute for Data Intensive Astronomy (IDIA). The computing equipment to run \simba\ was funded by BEIS capital funding via STFC capital grants ST/P002293/1, ST/R002371/1 and ST/S002502/1, Durham University and STFC operations grant ST/R000832/1. DiRAC is part of the National e-Infrastructure.  We acknowledge the use of computing facilities of IDIA for part of this work. IDIA is a partnership of the Universities of Cape Town, of the Western Cape and of Pretoria.

\section{Data availability}

The data underlying this article will be shared on reasonable request to the corresponding author. \simba\ snapshots and galaxy catalogs used for this work can be found at \url{http://simba.roe.ac.uk/}.





\footnotesize{
  \bibliographystyle{mnras}
  \bibliography{bibliography}
}




\bsp	
\label{lastpage}
\end{document}